\documentclass[12pt]{article}
\usepackage{graphicx}
\usepackage{multirow}
\usepackage{subfigure}
\input epsf

\usepackage{subfigure}

\usepackage{graphicx}
\usepackage{lscape}
\usepackage{citesort}
\usepackage{amssymb}
\usepackage{appendix}
\usepackage{multirow}
\usepackage[table]{xcolor}
\usepackage{colortbl}
\definecolor{lightgray}{gray}{0.9}
\usepackage{mathrsfs}
\begin{document}
\def\qq{\langle \bar q q \rangle}
\def\uu{\langle \bar u u \rangle}
\def\dd{\langle \bar d d \rangle}
\def\sp{\langle \bar s s \rangle}
\def\GG{\langle g_s^2 G^2 \rangle}
\def\Tr{\mbox{Tr}}
\def\figt#1#2#3{
        \begin{figure}
        $\left. \right.$
        \vspace*{-2cm}
        \begin{center}
        \includegraphics[width=10cm]{#1}
        \end{center}
        \vspace*{-0.2cm}
        \caption{#3}
        \label{#2}
        \end{figure}
    }

\def\figb#1#2#3{
        \begin{figure}
        $\left. \right.$
        \vspace*{-1cm}
        \begin{center}
        \includegraphics[width=10cm]{#1}
        \end{center}
        \vspace*{-0.2cm}
        \caption{#3}
        \label{#2}
        \end{figure}
                }

\def\ds{\displaystyle}
\def\beq{\begin{equation}}
\def\eeq{\end{equation}}
\def\bea{\begin{eqnarray}}
\def\eea{\end{eqnarray}}
\def\beeq{\begin{eqnarray}}
\def\eeeq{\end{eqnarray}}
\def\ve{\vert}
\def\vel{\left|}
\def\ver{\right|}
\def\nnb{\nonumber}
\def\ga{\left(}
\def\dr{\right)}
\def\aga{\left\{}
\def\adr{\right\}}
\def\lla{\left<}
\def\rra{\right>}
\def\rar{\rightarrow}
\def\lrar{\leftrightarrow}
\def\nnb{\nonumber}
\def\la{\langle}
\def\ra{\rangle}
\def\ba{\begin{array}}
\def\ea{\end{array}}
\def\tr{\mbox{Tr}}
\def\ssp{{\Sigma^{*+}}}
\def\sso{{\Sigma^{*0}}}
\def\ssm{{\Sigma^{*-}}}
\def\xis0{{\Xi^{*0}}}
\def\xism{{\Xi^{*-}}}
\def\qs{\la \bar s s \ra}
\def\qu{\la \bar u u \ra}
\def\qd{\la \bar d d \ra}
\def\qq{\la \bar q q \ra}
\def\gGgG{\la g^2 G^2 \ra}
\def\q{\gamma_5 \not\!q}
\def\x{\gamma_5 \not\!x}
\def\g5{\gamma_5}
\def\sb{S_Q^{cf}}
\def\sd{S_d^{be}}
\def\su{S_u^{ad}}
\def\sbp{{S}_Q^{'cf}}
\def\sdp{{S}_d^{'be}}
\def\sup{{S}_u^{'ad}}
\def\ssp{{S}_s^{'??}}

\def\sig{\sigma_{\mu \nu} \gamma_5 p^\mu q^\nu}
\def\fo{f_0(\frac{s_0}{M^2})}
\def\ffi{f_1(\frac{s_0}{M^2})}
\def\fii{f_2(\frac{s_0}{M^2})}
\def\O{{\cal O}}
\def\sl{{\Sigma^0 \Lambda}}
\def\es{\!\!\! &=& \!\!\!}
\def\ap{\!\!\! &\approx& \!\!\!}
\def\md{\!\!\!\! &\mid& \!\!\!\!}
\def\ar{&+& \!\!\!}
\def\ek{&-& \!\!\!}
\def\kek{\!\!\!&-& \!\!\!}
\def\cp{&\times& \!\!\!}
\def\se{\!\!\! &\simeq& \!\!\!}
\def\eqv{&\equiv& \!\!\!}
\def\kpm{&\pm& \!\!\!}
\def\kmp{&\mp& \!\!\!}
\def\mcdot{\!\cdot\!}
\def\erar{&\rightarrow&}
\def\olra{\stackrel{\leftrightarrow}}
\def\ola{\stackrel{\leftarrow}}
\def\ora{\stackrel{\rightarrow}}

\def\simlt{\stackrel{<}{{}_\sim}}
\def\simgt{\stackrel{>}{{}_\sim}}


\title{
         {\Large
                 {\bf
                      Thermal behavior of the mass and residue of hyperons  
                 }
         }
      }

\author{ K. Azizi \thanks {e-mail: kazizi@dogus.edu.tr}\,\,, G. Kaya \thanks {e-mail: gkaya@dogus.edu.tr}  \\
Department of Physics, Do\u gu\c s
University,
 Ac{\i}badem-Kad{\i}k\"oy, \\ 34722 Istanbul, Turkey\\
 }
\date{}

\begin{titlepage}
\maketitle
\thispagestyle{empty}

\begin{abstract}
We investigate the mass and residue of the  $\Sigma$, $\Lambda$ and $\Xi$ hyperons at finite temperature in the framework of thermal QCD sum rules. In our calculation, 
we take into account the additional operators coming up at finite temperature. We find the temperature-dependent continuum threshold for each hyperon using the obtained 
sum rules for their mass and residue. The numerical results demonstrate that the mass and residue of the particles under consideration remain stable up to a certain temperature, 
after which they decrease  by increasing the temperature.

\end{abstract}

~~~PACS number(s): 11.10.Wx, 11.55.Hx, 14.20.-c, 14.20.Jn
\end{titlepage}

\section{Introduction}


Investigation of the hadronic parameters at finite temperature is an essential tool in understanding the properties of hot and dense QCD matter produced by the heavy ion collision experiments. 
It can also provides us with useful knowledge on the internal structures of the dense astrophysical objects like neutron stars. Both in heavy ion collision experiments and
 at the center of the neutron stars where the density is high some strange particles like $\Sigma$, $\Lambda$ and $\Xi$ hyperons as well as some resonances may be  produced besides the nucleons. Hence, calculation
of many parameters related to such particles in hot medium can be useful in analysis of the results of  those experiments and the structure of dense stars. Besides, it is believed that the hadrons are melted near to a critical temperate
and another phase of matter called quark-gluon plasma (QGP) is appeared. Such phase of matter  may also be produced  both in  heavy collision experiments as well as in the core of neutron stars. Investigation of properties of
particle like hyperons near to the critical temperature can also help us in understanding of such new phase of matter.
  
To determine the in-medium hadronic parameters, 
some nonperturbative approaches are needed. One of the most applicable and powerful techniques among these methods is the QCD sum rule approach. The zero-temperature version of 
this method first introduced by Shifman, Vainshtein and Zakharov \cite{Shifman}, which has then been extended to finite temperature by Bochkarev and Shaposhnikov \cite{Bochkarev}. At finite temperature, some new operators appear in the operator product expansion (OPE) and the vacuum condensates are replaced by their thermal expectation values. At finite temperature, the Lorentz invariance is broken with the choice of the thermal rest frame and to restore it, the four-velocity vector of the medium is encountered.

In the literature, the properties of mesons at finite temperature have been extensively investigated within various phenomenological approaches, but there are a few studies devoted to the analysis of the nucleons and hyperons at finite temperature. The study of parameters of nucleons and hyperons ($\Sigma$, $\Lambda$ and $\Xi$) at finite temperature is very crucial for the understanding of the SU(3) flavor symmetry breaking. In \cite{Ryu}, the change of the magnetic moments and masses of octet baryons at a finite density and temperature has been investigated via quark-meson coupling (QMC) and modified quark-meson coupling (MQMC) models. It has been obtained that the magnetic moment of octet baryons increase with temperature, while their masses fall with increasing temperature. In \cite{Koike}, correlators of the octet baryons at finite temperature are studied in the framework of the QCD sum rule. In this work, it is demonstrated that the $O(T^{2})$ dependence of the condensates appearing in the OPE is totally absorbed by the scattering with the thermal pions and the change of the pole residues. In \cite{Burgio}, a parametrization of the free energy density of the hyper-nuclear matter at finite temperature is obtained within the Brueckner-Hartree-Fock (BHF) framework. In \cite{Rios}, the authors have studied the bulk and single-particle properties of the hot hyperonic matter within again the BHF approximation extended to finite temperature. It is found that the presence of hyperons can modify the thermodynamical properties of the system, considerably.


In this study, we extend our previous work on the nucleon properties at finite temperature \cite{Azizi} and use the QCD sum rule method to calculate the mass and residue 
of the $\Sigma$, $\Lambda$ and $\Xi$ hyperons at finite temperature using the most general form of their interpolating currents. We take into account the additional 
operators appearing in the OPE at finite temperature. In the calculations, we use the thermal light quark propagator in coordinate space containing both the perturbative 
and non-perturbative contributions. We also use the temperature-dependent quark and gluon condensates as well as the temperature-dependent fermionic and gluonic parts of the 
energy density to numerically analyze the sum rules for the quantities under consideration. We find the temperature-dependent expression of the continuum threshold for first
time in the baryonic channels for each hyperon using the obtained sum rules for the mass and residue. 

The paper is organized as follows. In section 2, the thermal QCD sum rules for the mass and residue of  $\Sigma$, $\Lambda$ and $\Xi$ hyperons are constructed. Section 3 is devoted to the numerical analyses of the obtained sum rules and comparison of the results with those existing in the literature.

\section{Thermal QCD sum rules for the mass and residue of $\Sigma$, $\Lambda$ and $\Xi$ hyperons}



To obtain the mass and residue of $\Sigma$, $\Lambda$ and $\Xi$ hyperons at finite temperature using the thermal QCD sum rules, we consider the following two-point thermal correlation function:

\begin{eqnarray}\label{Eq1}
\Pi(p,T)=i \int d^{4}x~ e^{ip\cdot x} Tr\Big(\rho~ {\cal
T}\Big(J_{h}(x)\bar{J}_{h}(0)\Big)\Big),
\end{eqnarray}
where $p$ stands for the four-momentum of the hyperon $h$, $\rho=e^{-\beta H}/Tr
e^{-\beta H}$ is the thermal density matrix of QCD at temperature
$T=1/\beta$ with $H$  being the QCD Hamiltonian and  $\cal T$ indicates the time
ordering operator. Also, $J_{h}(x)$ is the interpolating current for the hyperon and it is given by the following expressions for the members under consideration in terms of the light
 quarks fields \cite{Belyaev,Lee}:

\begin{equation}
J_{\Sigma}(x)=-\frac{1}{\sqrt{2•}}\epsilon_{abc}\sum_{i=1}^{2}\Bigg[\Big(u^{T,a}(x)CA_{1}^{i}s^{b}(x)\Big)
A_{2}^{i}d^{c}(x)-\Big(s^{T,a}(x)CA_{1}^{i}d^{b}(x)\Big)
A_{2}^{i}u^{c}(x)\Bigg],
\end{equation}
\begin{eqnarray}
J_{\Lambda}(x)&=&\frac{1}{\sqrt{6}}\epsilon_{abc}\sum_{i=1}^{2}\Bigg[2\Big(u^{T,a}(x)CA_{1}^{i}d^{b}(x)\Big)
A_{2}^{i}s^{c}(x)+\Big(u^{T,a}(x)CA_{1}^{i}s^{b}(x)\Big)
A_{2}^{i}d^{c}(x)\nonumber\\
&+&\Big(s^{T,a}(x)CA_{1}^{i}d^{b}(x)\Big)
A_{2}^{i}u^{c}(x)\Bigg],
\end{eqnarray}
and
\begin{equation}
J_{\Xi}(x)=-2\epsilon_{abc}\sum_{i=1}^{2}\Bigg[\Big(s^{T,a}(x)CA_{1}^{i}u^{b}(x)\Big)
A_{2}^{i}s^{c}(x)\Bigg],
\end{equation}
where $C$ is the charge conjugation
operator, $a, b, c$ are color indices; and $A_{1}^{1}=I$, $A_{1}^{2}=A_{2}^{1}=\gamma_5$ and 
$A_{2}^{2}=t$. In the above interpolating currents, $t$ is an arbitrary mixing parameter
with $t=-1$  being corresponding to the famous Ioffe current.

To proceed, we calculate the aforesaid correlation function in two sides, namely the hadronic and OPE. By matching the obtained results through a dispersion relation,
 we relate the hadronic parameters to the fundamental QCD degrees of freedom.

To obtain the correlation function in hadronic side,  we insert a  complete set of hyperonic state with spin $s$ into Eq. (1) and perform integral over four-$x$.  As a result  we get
\begin{eqnarray}\label{phepi}
\Pi^{Had}(p,T)=-\frac{{\langle}0|J_{h}(0)|h(p^{*},s){\rangle}_T
{\langle}h(p^{*}, s)|J_{h}^{\dag}(0)|0{\rangle}_T}{p^{*2}-m_{h}^{*2}}+...,
\end{eqnarray}
where label $T$ denotes that the matrix elements are calculated in the hot medium and   $...$ stands for the contributions of the higher states and  continuum. 
Here $p^*$ is the in-medium four-momentum of the corresponding hyperon and   $m^*_{h}$ is its modified mass. 
The matrix element  $\langle0|J_{h}(0)|h(p^{*},s)\rangle_T$ is defined as
\begin{eqnarray}\label{intcur}
\langle0|J_{h}(0)|H(p^{*},s)\rangle_T&=&\lambda_{h}(T)u_{h}(p^*,s) ,
\end{eqnarray}
where $\lambda_{h}(T)$ is the  temperature-dependent  residue of the corresponding hyperon and $u_{h}(p^*,s)$ is the Dirac spinor. By using Eq. (\ref{intcur})  in Eq. (\ref{phepi}) and summing over the spin, 
we find
\begin{equation}
\Pi^{Had}(p,T)=-\frac{\lambda_{h}^{2}(T)[\!\not\!{p^*}+m^*_{h}]}{p^{*2}-m_{h}^{*2}}+... ,
\end{equation}
which can be written as
\begin{eqnarray}\label{jg}
\Pi^{Had}(p,T)&=&-\frac{\lambda_{h}^{2}(T)}{\!\not\!p^{*}-m^*_{h}}+..., \nonumber \\
&=&-\frac{\lambda_{h}^{2}(T)}{
(p^{\mu}-\sigma u^{\mu})\gamma_\mu-m^*_{h}} +... . \nonumber \\
\end{eqnarray}
where $u^{\mu}$ is the four-velocity vector of the medium  and $\sigma$ is the self-energy  at finite temperature. We shall work in the rest frame of the heat bath, i.e. $u_{\mu}=(1,0,0,0)$ implying
that $u^2=1$ and $p\cdot u=p_{0}$. In this frame, we can write the second line of  Eq. (\ref{jg}) as
\begin{eqnarray}
 \Pi^{Had}(p,T)&=&-\frac{\lambda_{h}^{2}(T)[\!\not\!{p}-\sigma\!\not\!{u} +m^*_{h}]}{p^{2}+\sigma^2-2\sigma p_0-m_{h}^{*2}}+...,\nonumber\\
&=&-\frac{\lambda_{h}^{2}(T)[\!\not\!{p}-\sigma\!\not\!{u} +m^*_{h}]}{p^{2}-m_{h}^{2}(T)}+...,
\end{eqnarray}
where $m_{h}^{2}(T)=m_{h}^{*2}-\Sigma^2+2\sigma p_0$ with $m_{h}(T)$ being the temperature-dependent mass of the hyperons under consideration. By considering  $\sigma\approx2p_0$ the correlation function at hadronic side can be written in terms of different structures as
\begin{eqnarray}\label{hadstruction}
\Pi^{Had}(p,T)= \Pi^{Had}_{p}(p^2,p_0,T)\!\not\!{p}+\Pi^{Had}_{u}(p^2,p_0,T)\!\not\!{u}+\Pi^{Had}_{S}(p^2,p_0,T)I, 
\end{eqnarray}
where $I$ stands for the unit matrix; and 
\begin{eqnarray}\label{correpslash1}
\Pi^{Had}_p(p^2,p_0,T)&=&-\lambda_h^{2}(T)\frac{1}{p^2-m_{h}^{2}(T)},
\end{eqnarray}
\begin{eqnarray}\label{correpslash1}
\Pi^{Had}_u(p^2,p_0,T)&=&\lambda_h^{2}(T)\frac{2p_0}{p^2-m_{h}^{2}(T)},
\end{eqnarray}
and
\begin{eqnarray}\label{correpslash2}
\Pi^{Had}_S(p^2,p_0,T)&=&-\lambda_h^{2}(T)\frac{m_{h}(T)}{p^2-m_{h}^{2}(T)}.
\end{eqnarray}
We will choose the structures $\not\!p$ and $I$  to calculate the mass and residue of the considered hyperons.
By applying the Borel 
transformation with respect to $p^2$ to suppress the contributions of the higher states and continuum, for the coefficient of the selected structures, we find 
\begin{eqnarray}
\hat{B}\Pi^{Had}_p(p_0,T)=-\lambda_h^{2}(T)e^{-m_{h}^{2}(T)/M^2},
\end{eqnarray}
and
\begin{eqnarray}
\hat{B}\Pi^{Had}_S(p_0,T)=-\lambda_h^{2}(T)m_{h}(T) e^{-m_{h}^{2}(T)/M^2},
\end{eqnarray}
 where $M^2$  is the Borel parameter.

Now, we calculate the OPE side of the thermal correlation function in deep Euclidean region. Similar to Eq. (\ref{hadstruction}), 
the OPE  side of the thermal correlation function can be written in terms of three distinct structures as
\begin{eqnarray} 
\Pi^{OPE}(p,T)=\Pi_{p}^{OPE}(p^2,p_0,T)\!\not\!{p}+\Pi_{u}^{OPE}(p^2,p_0,T)\!\not\!{u}+\Pi_{S}^{OPE}(p^2,p_0,T)I.\nonumber \\
\end{eqnarray}
Our main task in the following is to calculate the functions $\Pi_{p}^{OPE}(p^2,p_0,T)$ and $\Pi_{S}^{OPE}(p^2,p_0,T)$. To this aim, we insert the explicit forms of the interpolating currents to the thermal correlation function and contract all quark pairs to obtain the result in terms of the light quarks propagators. We get the following expressions for the 
correlation function of the considered hyperons in OPE side in $x$ space:

\begin{eqnarray}\label{corre2}
\Pi_{\Sigma}^{OPE}(p,T) &=& \frac{i}{2}\epsilon_{abc}\epsilon_{a'b'c'}\int d^4 x
e^{ipx}\Bigg\langle 
\Bigg\{\Bigg(\gamma_{5}S^{cc'}_{d}(x)\gamma_{5}Tr\Bigg[S^{ab'}_{u}(x)S'^{ba'}_{s}(x)\Bigg]\nonumber \\
&+&\gamma_{5}S^{ca'}_{d}(x)S'^{bb'}_{s}(x)S^{ac'}_{u}(x)\gamma_{5}+\gamma_{5}S^{cb'}_{u}(x)S'^{aa'}_{s}(x)S^{bc'}_{d}(x)\gamma_{5}\nonumber \\
&+&\gamma_{5}S^{cc'}_{u}(x)\gamma_{5}Tr\Bigg[S^{ab'}_{s}(x)S'^{ba'}_{d}(x)\Bigg]\Bigg)
+t\Bigg(\gamma_{5}S^{ca'}_{d}(x)\gamma_{5}S'^{bb'}_{s}(x)S^{ac'}_{u}(x)\nonumber \\
&+&\gamma_{5}S^{cc'}_{d}(x)Tr\Bigg[\gamma_{5}S'^{ba'}_{s}(x)S^{ab'}_{u}(x)\Bigg]+S^{cb'}_{u}(x)S'^{aa'}_{s}(x)\gamma_{5}S^{bc'}_{d}(x)\gamma_{5}\nonumber \\
&+&S^{cc'}_{u}(x)\gamma_{5}Tr\Bigg[S'^{ba'}_{d}(x)\gamma_{5}S^{ab'}_{s}(x)\Bigg]+\gamma_{5}S^{cc'}_{u}(x)Tr\Bigg[\gamma_{5}S'^{ba'}_{d}(x)S^{ab'}_{s}(x)\Bigg]\nonumber \\
&+&\gamma_{5}S^{cb'}_{u}(x)\gamma_{5}S'^{aa'}_{s}(x)S^{bc'}_{d}(x)+S^{cc'}_{d}(x)\gamma_{5}Tr\Bigg[S'^{ba'}_{s}(x)\gamma_{5}S^{ab'}_{u}(x)\Bigg]\nonumber \\
&+&S^{ca'}_{d}(x)S'^{bb'}_{s}(x)\gamma_{5}S^{ac'}_{u}(x)\gamma_{5}\Bigg)+t^2\Bigg(S^{cc'}_{u}(x)Tr\Bigg[\gamma_{5}S'^{ab'}_{s}(x)\gamma_{5}S^{ba'}_{d}(x)\Bigg]\nonumber \\
&+&S^{cb'}_{u}(x)\gamma_{5}S'^{aa'}_{s}(x)\gamma_{5}S^{bc'}_{d}(x)+S^{ca'}_{d}(x)\gamma_{5}S'^{bb'}_{s}(x)\gamma_{5}S^{ac'}_{u}(x)\nonumber \\
&+&S^{cc'}_{d}(x)Tr\Bigg[\gamma_{5}S'^{ab'}_{u}(x)\gamma_{5}S^{ba'}_{s}(x)\Bigg]\Bigg) \Bigg\} \Bigg\rangle_{T},
\end{eqnarray}
\begin{eqnarray}\label{corre3}
\Pi_{\Lambda}^{OPE}(p,T) &=& \frac{i}{6}\epsilon_{abc}\epsilon_{a'b'c'}\int d^4 x
e^{ipx}\Bigg\langle 
\Bigg\{\Bigg(4\gamma_{5}S^{cc'}_{s}(x)\gamma_{5}Tr\Bigg[S^{ab'}_{u}(x)S'^{ba'}_{d}(x)\Bigg]\nonumber \\
&-&2\gamma_{5}S^{ca'}_{s}(x)S'^{ab'}_{u}(x)S^{bc'}_{d}(x)\gamma_{5}-2\gamma_{5}S^{cb'}_{s}(x)S'^{ba'}_{d}(x)S^{ac'}_{u}(x)\gamma_{5}\nonumber \\
&-&2\gamma_{5}S^{ca'}_{d}(x)S'^{ab'}_{u}(x)S^{bc'}_{s}(x)\gamma_{5}+\gamma_{5}S^{cc'}_{d}(x)\gamma_{5}Tr\Bigg[S'^{ba'}_{s}(x)S^{ab'}_{u}(x)\Bigg]\nonumber \\
&-&\gamma_{5}S^{ca'}_{d}(x)S'^{bb'}_{s}(x)S^{ac'}_{u}(x)\gamma_{5}-2\gamma_{5}S^{cb'}_{u}(x)S'^{ba'}_{d}(x)S^{ac'}_{s}(x)\gamma_{5}\nonumber \\
&-&\gamma_{5}S^{cb'}_{u}(x)S'^{aa'}_{s}(x)S^{bc'}_{d}(x)\gamma_{5}+\gamma_{5}S^{cc'}_{u}(x)\gamma_{5}Tr\Bigg[S'^{ba'}_{d}(x)S^{ab'}_{s}(x)\Bigg]\Bigg)\nonumber \\
&+&t\Bigg(4\gamma_{5}S^{cc'}_{s}(x)Tr\Bigg[\gamma_{5}S'^{ba'}_{d}(x)S^{ab'}_{u}(x)\Bigg]-2\gamma_{5}S^{ca'}_{s}(x)\gamma_{5}S'^{ab'}_{u}(x)S^{bc'}_{d}(x)\nonumber \\
&-&2\gamma_{5}S^{cb'}_{s}(x)\gamma_{5}S'^{ba'}_{d}(x)S^{ac'}_{u}(x)+4S^{cc'}_{s}(x)\gamma_{5}Tr\Bigg[S'^{ba'}_{d}(x)\gamma_{5}S^{ab'}_{u}(x)\Bigg]\nonumber \\
&-&2S^{ca'}_{s}(x)S'^{ab'}_{u}(x)\gamma_{5}S^{bc'}_{d}(x)\gamma_{5}-2S^{cb'}_{s}(x)S'^{ba'}_{d}(x)\gamma_{5}S^{ac'}_{u}(x)\gamma_{5}\nonumber \\
&-&2\gamma_{5}S^{ca'}_{d}(x)\gamma_{5}S'^{ab'}_{u}(x)S^{bc'}_{s}(x)+\gamma_{5}S^{cc'}_{d}(x)Tr\Bigg[\gamma_{5}S'^{ba'}_{s}(x)S^{ab'}_{u}(x)\Bigg]\nonumber \\
&-&\gamma_{5}S^{ca'}_{d}(x)\gamma_{5}S'^{bb'}_{s}(x)S^{ac'}_{u}(x)-2S^{ca'}_{d}(x)S'^{ab'}_{u}(x)\gamma_{5}S^{bc'}_{s}(x)\gamma_{5}\nonumber \\
&+&S^{cc'}_{d}(x)\gamma_{5}Tr\Bigg[S'^{ba'}_{s}(x)\gamma_{5}S^{ab'}_{u}(x)\Bigg]-S^{ca'}_{d}(x)S'^{bb'}_{s}(x)\gamma_{5}S^{ac'}_{u}(x)\gamma_{5}\nonumber \\
&-&2\gamma_{5}S^{cb'}_{u}(x)\gamma_{5}S'^{ba'}_{d}(x)S^{ac'}_{s}(x)-\gamma_{5}S^{cb'}_{u}(x)\gamma_{5}S'^{aa'}_{s}(x)S^{bc'}_{d}(x)\nonumber \\
&+&\gamma_{5}S^{cc'}_{u}(x)Tr\Bigg[\gamma_{5}S'^{ba'}_{d}(x)S^{ab'}_{s}(x)\Bigg]-2S^{cb'}_{u}(x)S'^{ba'}_{d}(x)\gamma_{5}S^{ac'}_{s}(x)\gamma_{5}\nonumber \\
&-&S^{cb'}_{u}(x)S'^{aa'}_{s}(x)\gamma_{5}S^{bc'}_{d}(x)\gamma_{5}+S^{cc'}_{u}(x)\gamma_{5}Tr\Bigg[S'^{ba'}_{d}(x)\gamma_{5}S^{ab'}_{s}(x)\Bigg]\Bigg)\nonumber \\
&+&t^2\Bigg(4S^{cc'}_{s}(x)Tr\Bigg[\gamma_{5}S'^{ab'}_{u}(x)\gamma_{5}S^{ba'}_{d}(x)\Bigg]-2S^{ca'}_{s}(x)\gamma_{5}S'^{ab'}_{u}(x)\gamma_{5}S^{bc'}_{d}(x)\nonumber \\
&-&2S^{cb'}_{s}(x)\gamma_{5}S'^{ba'}_{d}(x)\gamma_{5}S^{ac'}_{u}(x)-2S^{ca'}_{d}(x)\gamma_{5}S'^{ab'}_{u}(x)\gamma_{5}S^{bc'}_{s}(x)\nonumber \\
&+&S^{cc'}_{d}(x)Tr\Bigg[\gamma_{5}S'^{ab'}_{u}(x)\gamma_{5}S^{ba'}_{s}(x)\Bigg]-S^{ca'}_{d}(x)\gamma_{5}S'^{bb'}_{s}(x)\gamma_{5}S^{ac'}_{u}(x)\nonumber \\
&-&2S^{cb'}_{u}(x)\gamma_{5}S'^{ba'}_{d}(x)\gamma_{5}S^{ac'}_{s}(x)-S^{cb'}_{u}(x)\gamma_{5}S'^{aa'}_{s}(x)\gamma_{5}S^{bc'}_{d}(x)\nonumber \\
&+&S^{cc'}_{u}(x)Tr\Bigg[\gamma_{5}S'^{ab'}_{s}(x)\gamma_{5}S^{ba'}_{d}(x)\Bigg]\Bigg) \Bigg\} \Bigg\rangle_{T},
\end{eqnarray}
and
\begin{eqnarray}\label{corre4}
\Pi_{\Xi}^{OPE}(p,T) &=& i\epsilon_{abc}\epsilon_{a'b'c'}\int d^4 x
e^{ipx}\Bigg\langle 
\Bigg\{\Bigg(\gamma_{5}S^{cc'}_{s}(x)\gamma_{5}Tr\Bigg[S^{ab'}_{s}(x)S'^{ba'}_{u}(x)\Bigg]\nonumber \\
&-&\gamma_{5}S^{cb'}_{s}(x)S'^{ba'}_{u}(x)S^{ac'}_{s}(x)\gamma_{5}\Bigg)+t\Bigg(\gamma_{5}S^{cc'}_{s}(x)Tr\Bigg[\gamma_{5}S'^{ba'}_{u}(x)S^{ab'}_{s}(x)\Bigg]\nonumber \\
&-&\gamma_{5}S^{cb'}_{s}(x)\gamma_{5}S'^{ba'}_{u}(x)S^{ac'}_{s}(x)+S^{cc'}_{s}(x)\gamma_{5}Tr\Bigg[S'^{ba'}_{u}(x)\gamma_{5}S^{ab'}_{s}(x)\Bigg]\nonumber \\
&-&S^{cb'}_{s}(x)S'^{ba'}_{u}(x)\gamma_{5}S^{ac'}_{s}(x)\gamma_{5}\Bigg)+t^2\Bigg(S^{cc'}_{s}(x)Tr\Bigg[\gamma_{5}S'^{ab'}_{s}(x)\gamma_{5}S^{ba'}_{u}(x)\Bigg]\nonumber \\
&-&S^{cb'}_{s}(x)\gamma_{5}S'^{ba'}_{u}(x)\gamma_{5}S^{ac'}_{s}(x)\Bigg) \Bigg\} \Bigg\rangle_{T},
\end{eqnarray}
where $S'=CS^TC$ and $S_{u,d,s}(x)$ denotes the thermal light quark propagator in $x$ space. The light quark propagator in vacuum is calculated in \cite{Reinders} in momentum space. 
One can easily transform the expression of the light-quark propagator in momentum space to the coordinate space via Fourier transformation (see for instance \cite{Wang}). As we also stated
in the introduction,  to restore the Lorentz invariance at finite temperature broken with the choice of the thermal rest frame where matter is
at rest at a definite temperature,  the four-velocity vector of the medium should be encountered.  In this condition, the residual O(3) invariance brings some extra operators with the same mass 
dimension as the vacuum condensates \cite{Mallik1,Weldon}. These new operators are obtained in \cite{Mallik}. Considering these extra operators, we can write  the thermal light quark propagator in
 coordinate space as (see also \cite{Veliev})
\begin{eqnarray}\label{lightquarkpropagator}
S_{q}^{ij}(x)&=& i\frac{\!\not\!{x}}{ 2\pi^2 x^4}\delta_{ij}
-\frac{m_q}{4\pi^2 x^2}\delta_{ij}-\frac{\langle
\bar{q}q\rangle}{12}\delta_{ij} -\frac{ x^{2}}{192} m_{0}^{2}
\langle
\bar{q}q\rangle\Big[1-i\frac{m_q}{6}\!\not\!{x}\Big]\delta_{ij}
\nonumber\\
&+&\frac{i}{3}\Big[\!\not\!{x}\Big(\frac{m_q}{16}\langle
\bar{q}q\rangle-\frac{1}{12}\langle u\Theta^{f}u\rangle\Big)
+\frac{1}{3}\Big(u\cdot x\!\not\!{u}\langle
u\Theta^{f}u\rangle\Big)\Big]\delta_{ij}
\nonumber\\
&-&\frac{ig_s \lambda_{A}^{ij}}{32\pi^{2} x^{2}}
G_{\mu\nu}^{A}\Big(\!\not\!{x}\sigma^{\mu\nu}+\sigma^{\mu\nu}
\!\not\!{x}\Big),
\end{eqnarray}
where $m_{q}$ is the light quark mass, $\langle\bar{q}q\rangle$ denotes the
temperature-dependent light quark condensate, $G_{\mu\nu}^{A}$ is the temperature-dependent external gluon field, 
$\Theta^{f}_{\mu\nu}$ is 
the fermionic part of the energy momentum tensor and $\lambda_{A}^{ij}$
are the Gell-Mann matrices. In principle, the light quark propagator  
contains also terms with Logarithms. However, we neglect these terms  since they are proportional to the light-quark mass and give  minimal contributions to sum rules. We also neglect the
 terms containing the four-quark operators as also give small contributions to the calculations.

The next step is to use the thermal light quark propagator in Eqs. (\ref{corre2}), (\ref{corre3}) and (\ref{corre4}) and perform Fourier integral to go to the momentum space. 
After applying the Borel transformation as well as the continuum subtraction we get the functions $\Pi_{p}^{OPE}(p_0,T)$ and $\Pi_{S}^{OPE}(p_0,T)$, containing both the perturbative and
non-perturbative contributions, for the hyperons under consideration as presented in the 
Appendix A. Note that for separation of the perturbative (short-distance effects) and non-perturbative (long-distance effects) we use a  factorization scale between $0.5~GeV$ and $1.0~GeV$
 (see also \cite{Chetyrkin,Balitsky}).  We also use the relation
\begin{eqnarray}\label{TrGG} 
\langle Tr^c G_{\alpha \beta} G_{\mu \nu}\rangle &=& \frac{1}{24} (g_{\alpha \mu} g_{\beta \nu} -g_{\alpha
\nu} g_{\beta \mu})\langle G^a_{\lambda \sigma} G^{a \lambda \sigma}\rangle \nonumber \\
 &+&\frac{1}{6}\Big[g_{\alpha \mu}
g_{\beta \nu} -g_{\alpha \nu} g_{\beta \mu}-2(u_{\alpha} u_{\mu}
g_{\beta \nu} -u_{\alpha} u_{\nu} g_{\beta \mu} -u_{\beta} u_{\mu}
g_{\alpha \nu} +u_{\beta} u_{\nu} g_{\alpha \mu})\Big]\nonumber \\
&\times&\langle u^{\lambda} {\Theta}^g _{\lambda \sigma} u^{\sigma}\rangle
\end{eqnarray}
 to express the two-gluon condensate in terms of the gluonic part of the energy-momentum tensor $\Theta^{g}_{\lambda \sigma}$. 

After matching the coefficients of the selected structures from the hadronic and OPE sides, the following thermal sum rules for the hyperons in Borel scheme are obtained:

\begin{eqnarray}\label{residuesumrule}
-\lambda_h^{2}(T)e^{-m_{h}^2(T)/M^2}=\hat{B}\Pi_{p,h}^{OPE}(p_0,T),
\end{eqnarray}
and
\begin{eqnarray}\label{residuesumrule2}
-\lambda_h^{2}(T)m_{h}(T)e^{-m_{h}^2(T)/M^2}=\hat{B}\Pi_{S,h}^{OPE}(p_0,T),
\end{eqnarray}
where the mass of the corresponding hyperon is obtained as
\begin{eqnarray} \label{ratio}
m_{h}(T)=\frac{\hat{B}\Pi_{S,h}^{OPE}(p_0,T)}{\hat{B}\Pi_{p,h}^{OPE}(p_0,T)}.
\end{eqnarray}

\section{Numerical results}

\begin{figure}[h!]
\begin{center}
\subfigure[]{\includegraphics[width=8cm]{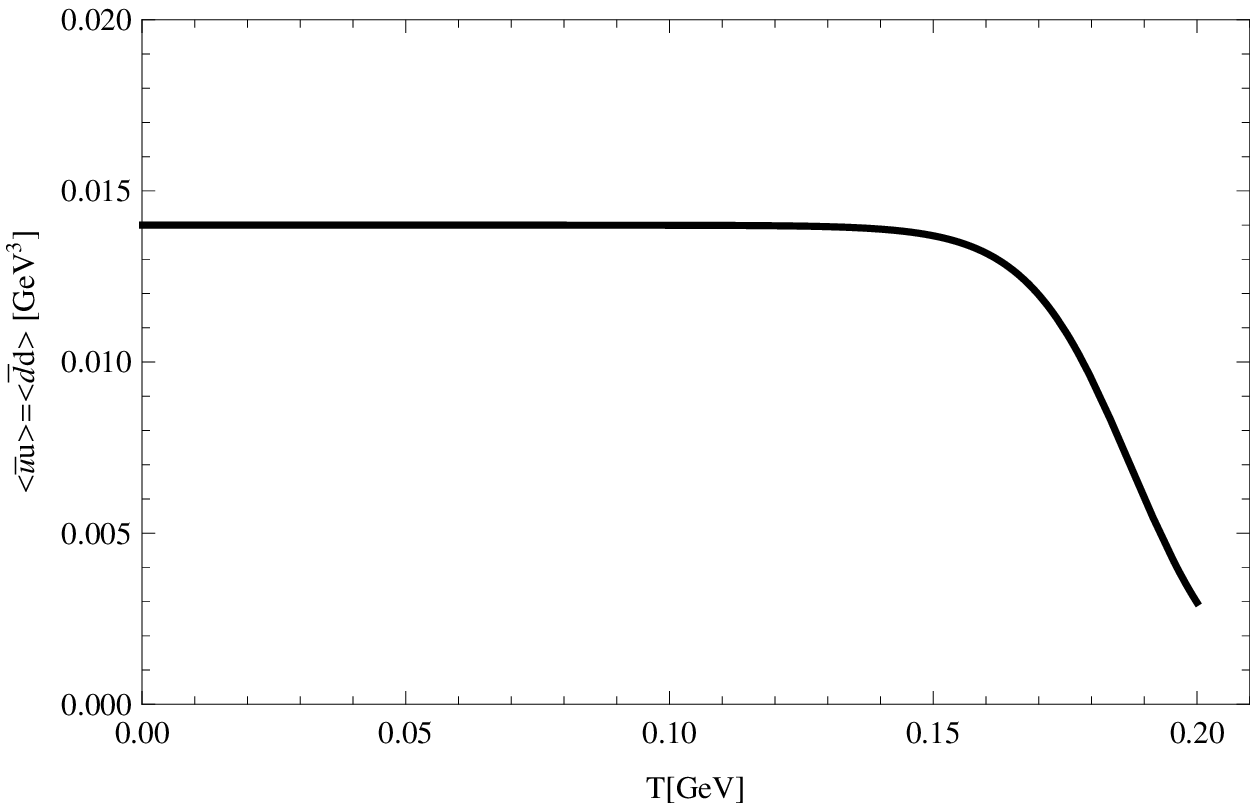}}
\subfigure[]{\includegraphics[width=8cm]{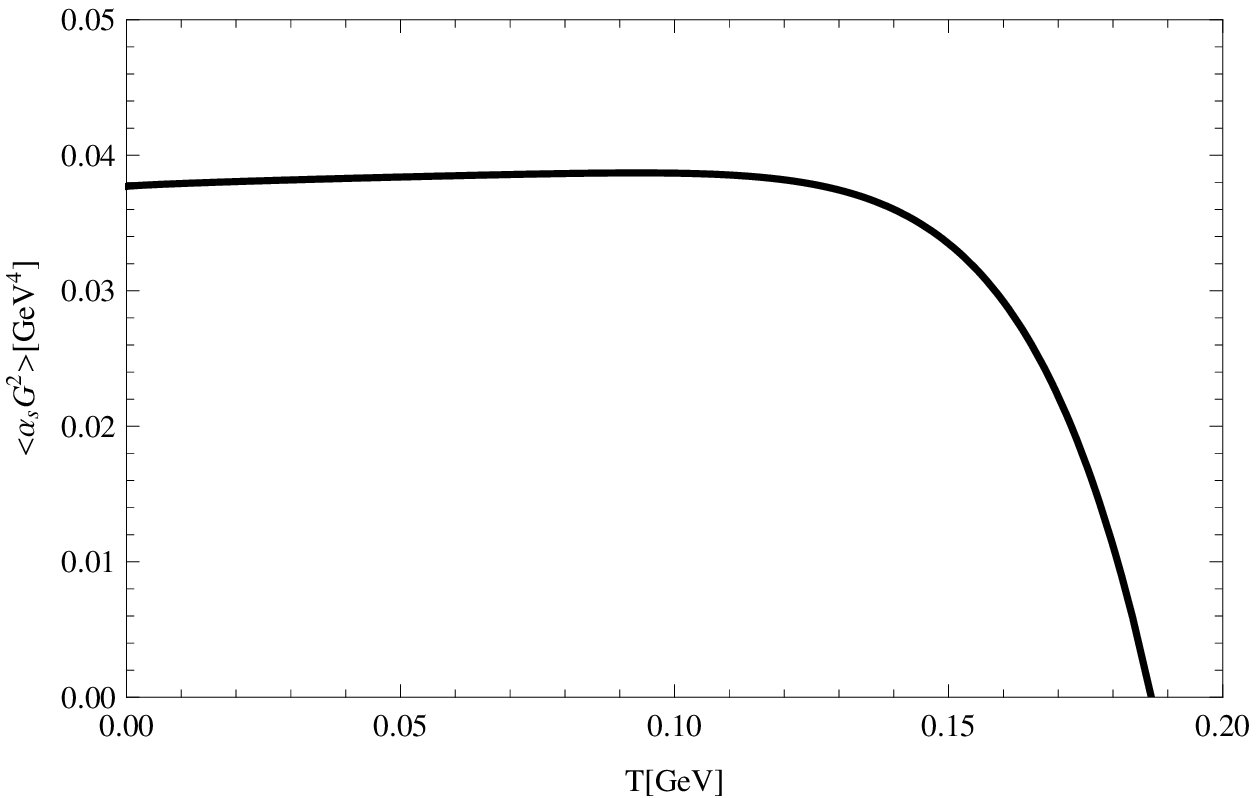}}
\subfigure[]{\includegraphics[width=8cm]{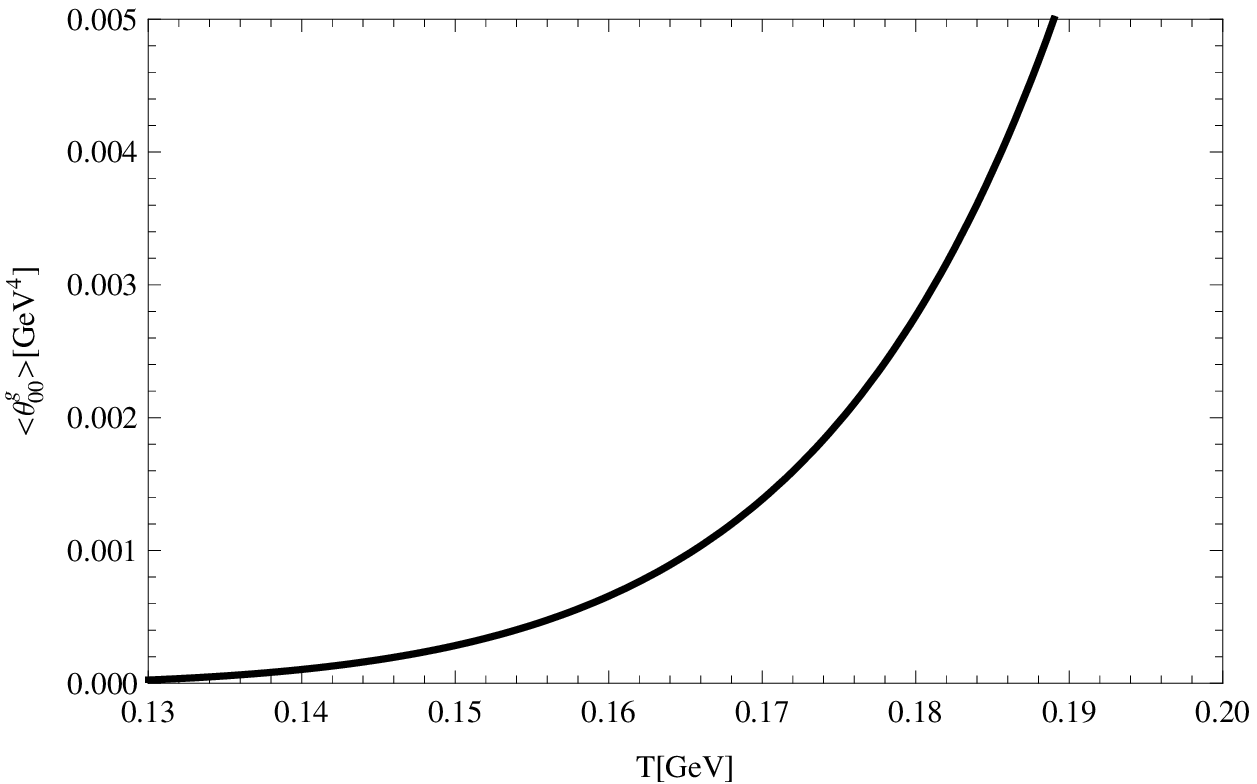}}
\end{center}
\caption{Variations of the quark condensate, gluon condensate and the energy density with respect to temperature. } \label{condd}
\end{figure}

This section is dedicated to the numerical analysis of the mass and residue of $\Sigma$, $\Lambda$ and $\Xi$ hyperons at finite temperature.
 We present the values of some input parameters used in the calculations in table \ref{table1}. To proceed, we also need to know the temperature-dependent quark and gluon condensates as well as the 
temperature-dependent energy density and continuum threshold. There are many calculations on the temperature-dependent condensates and energy-momentum tensor; however, we use the most recent and reliable
 results obtained via lattice QCD as well as QCD sum rules. In the case of quark condensate and energy density, we find  fit functions, which describe well the graphics on the variations of 
these parameters with respect 
to temperature
presented in Refs. \cite{Ayala,Bazavov,Cheng1,M.Cheng}. For the temperature-dependent gluon condensate we directly use the parametrization also obtained via QCD sum rules and data from lattice
 QCD in Ref. \cite{Ayala2}. These parameterizations
reproduce well the variations of different observables in some hadronic channels with respect to temperature (see Refs. \cite{Ayala,Bazavov,Cheng1,M.Cheng,Ayala2} for details). 
\begin{table}[ht!]
\centering
\rowcolors{1}{lightgray}{white}
\begin{tabular}{cc}
\hline \hline
   Parameters  &  Values    
           \\
\hline \hline
$p_0  $   (for  $\Sigma$ hyperon)       &  $1.192~GeV$     \\
$p_0  $   (for   $\Lambda$ hyperon)       &  $1.115~GeV$     \\
$p_0  $   (for  $\Xi$ hyperon)       &  $1.314~GeV$     \\
$ m_{u}   $          &  $(2.3_{-0.5}^{+0.7})$ $MeV$      \\
$ m_{d}   $          &   $(4.8_{-0.3}^{+0.5})$ $MeV$    \\
$ m_{s}   $          &   $(95\pm5 )$ $MeV$    \\
$ m_{0}^{2}   $          &  $(0.8\pm0.2)$ $GeV^2$         \\
$ \langle0|\overline{u}u|0\rangle = \langle0|\overline{d}d|0\rangle$          &  $-(0.24\pm0.01)^3$ $GeV^3$       \\
$ \langle0|\overline{s}s|0\rangle $          &  $-0.8(0.24\pm0.01)^3$ $GeV^3$       \\
$ {\langle}0\mid \frac{1}{\pi}\alpha_s G^2 \mid 0{\rangle}$          &  $ (0.012\pm0.004)~GeV^4$   \\
\hline \hline
\end{tabular}
\caption{The values of some input parameters used in numerical calculations \cite{Olive,H.G.Dosch,Belyaev,B.L.Ioffe}. } \label{table1}
\end{table}
To this end, for the  temperature-dependent quark condensate we obtain the fit function 
\begin{eqnarray}\label{qbarq}
\langle\bar{q}q\rangle=\frac{\langle0|\bar{q}q|0\rangle}{1+e^{18.10042(1.84692[\frac{1}{GeV^2}] T^{2}+4.99216[\frac{1}{GeV}] T-1)}},
\end{eqnarray}
where $\langle0|\bar{q}q|0\rangle$ stands for the vacuum light-quark condensate. This function is valid up to a critical temperature $ T_{c}=197~MeV$ and reproduce  the lattice QCD  and QCD sum rules 
results presented in 
\cite{Ayala,Bazavov,Cheng1}.  For the temperature-dependent gluon condensate we use the following parametrization obtained via QCD sum rules predictions and lattice QCD data  in  \cite{Ayala2}:
\begin{eqnarray}\label{G2TLattice}
\langle G^{2}\rangle=\langle 0|G^{2}|0\rangle\Bigg[1-1.65\Big(\frac{T}{T_{c}}\Big)^{8.735}+0.04967\Big(\frac{T}{T_{c}}\Big)^{0.7211}\Bigg].
\end{eqnarray}
Finally, we find and use the following parametrization for the gluonic and fermionic parts of the energy density using the lattice QCD graphics presented in \cite{M.Cheng}:
\begin{eqnarray}\label{tetamumu}
\langle\Theta^{g}_{00}\rangle=\langle\Theta^{f}_{00}\rangle=T^{4} e^{(113.867[\frac{1}{GeV^2}] T^{2}-12.190[\frac{1}{GeV}] T)}-10.141 [\frac{1}{GeV}]T^{5},
\end{eqnarray}
where this parametrization is valid for $T\geq130~MeV$. Note that we take the gluonic part of the energy density equal to its fermionic part since the contributions of terms containing
these two components is very small (let say $4\%$ of the total contribution roughly at any temperature). The variations of the quark condensate, gluon condensate and the energy density with respect to temperature is shown in figure
\ref{condd}.

In this study, we find the temperature-dependent continuum threshold for first time in baryonic channels for each hyperon using the obtained sum 
rules for the mass and residue in Eqs.  (\ref{residuesumrule}) and  (\ref{residuesumrule2}) as well as an extra equation obtained from Eq.  (\ref{residuesumrule}) by applying a 
derivative with respect to $-\frac{1}{M^2}$ to both sides. By eliminating the mass and residue from these equations, we found that the temperature-dependent continuum threshold can be parametrized as

\begin{eqnarray}\label{continuumthreshold}
s_{0}(T)=s_{0}\Bigg[1-\Big(\frac{T}{T_{c}}\Big)^{12}\Bigg],
\end{eqnarray}
where $s_{0}$ is the continuum threshold in vacuum.
This parameter is not totally arbitrary, but it depends on the energy of the first excited state with the same quantum numbers as the baryons under consideration. We take the interval
$[m_{h}(0)+0.4]^{2}~GeV^2\leq s_0\leq[m_{h}(0)+0.6]^{2}~GeV^2$ for this parameter, with $m_{h}(0)$  being the hyperon mass at $T=0$, at which the dependence of the physical observables on
 this parameter is weak according to our calculations.
%
%
%
%



Next, we try to determinate the working regions for two auxiliary parameters, namely the Borel mass parameter $M^{2}$ and the mixing parameter $t$, such that at these regions the physical quantities are
 roughly independent of these auxiliary parameters. For determination of the working region for the Borel parameter we require  that not only the contributions of the higher states and continuum 
should adequately be suppressed, but also the perturbative part should exceed the non-perturbative contributions and the series of the OPE converge. In technique language, 
 the upper bound on Borel parameter is obtained by demanding that 

\begin{eqnarray}
\label{nolabel}
\frac{ \int_{0}^{s_0}ds  \rho_{p,S}(s) e^{-s/M^2} }{
 \int_{0}^\infty ds \rho_{p,S}(s) e^{-s/M^2}} ~~>~~ 1/2, 
\end{eqnarray}
where $\rho_{p,S}(s)$ are the spectral densities corresponding to different structures for the corresponding hyperons. The lower bound on $M^{2}$ is determined by requiring that the perturbative contribution
exceeds that of the non-perturbative; and the term with higher dimension constitutes less than 10\% of the total contribution.

\begin{figure}[h!]
\begin{center}
\subfigure[]{\includegraphics[width=8cm]{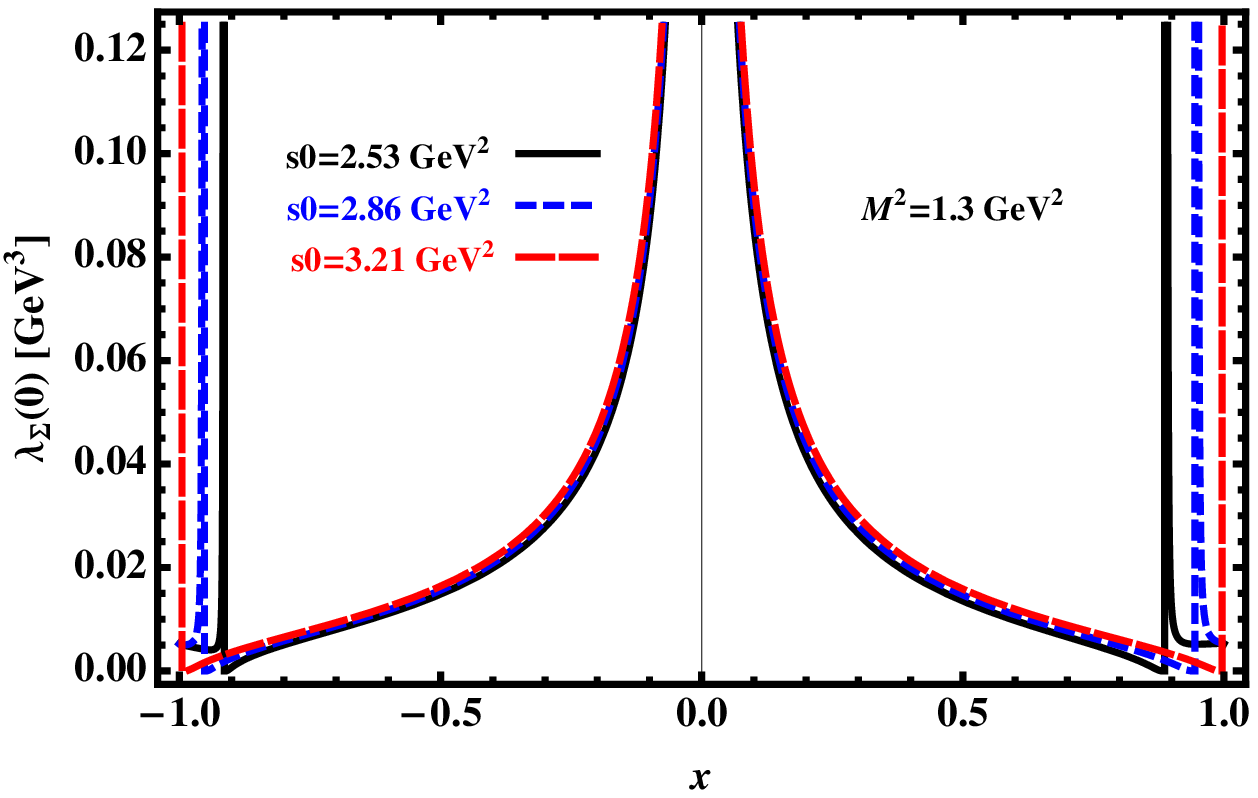}}
\subfigure[]{\includegraphics[width=8cm]{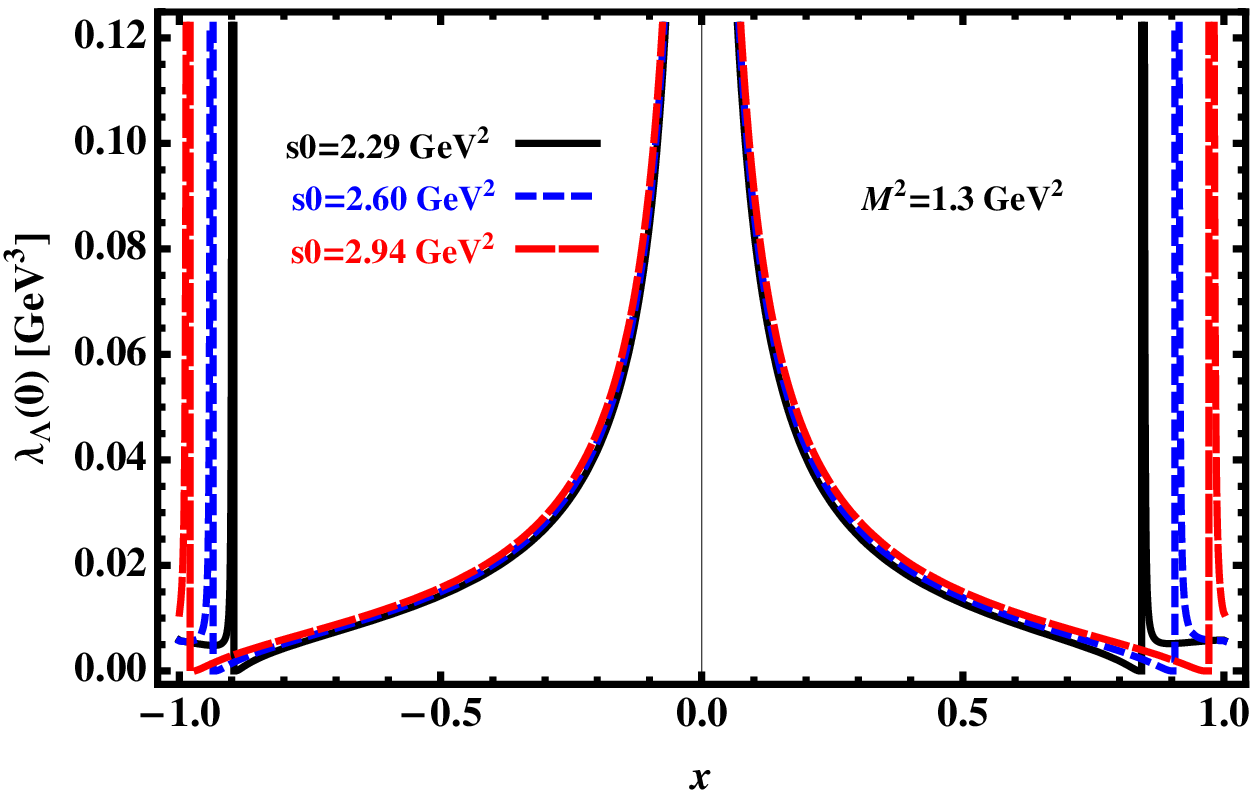}}
\subfigure[]{\includegraphics[width=8cm]{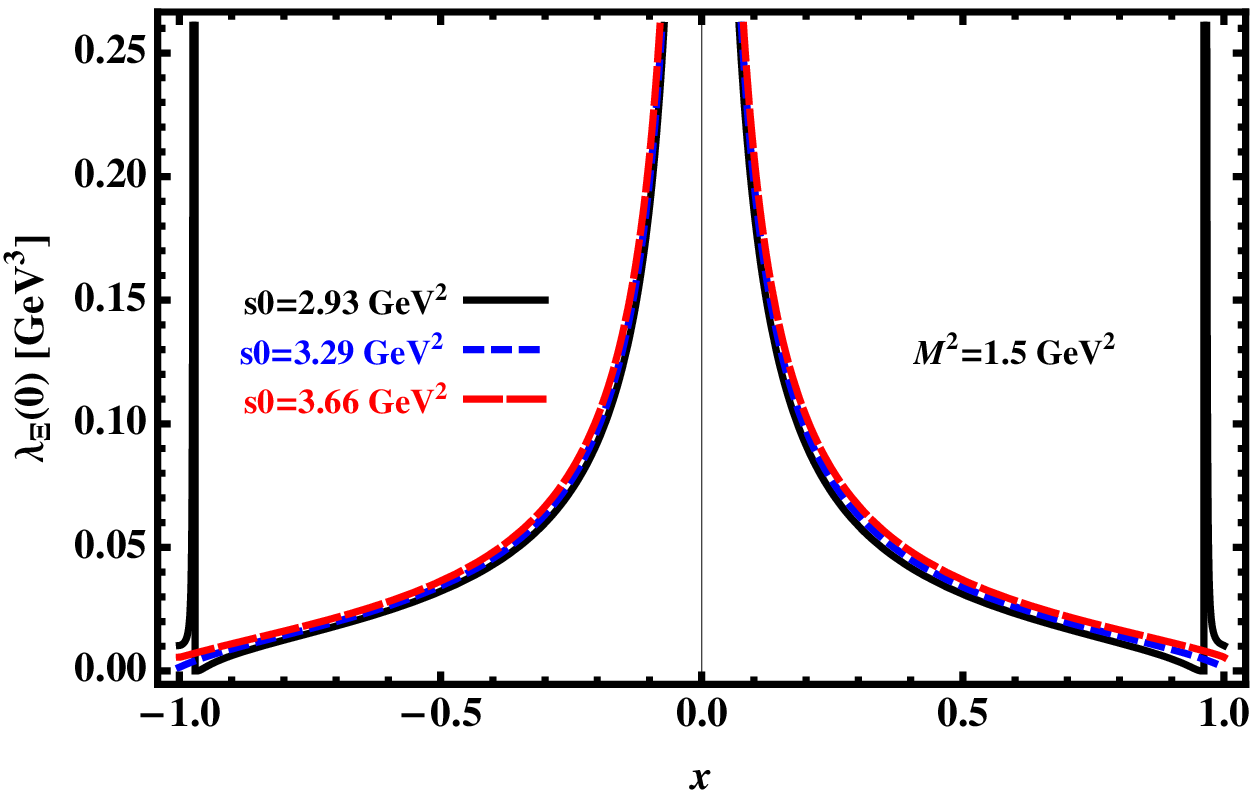}}
\end{center}
\caption{(a) The residue of the $\Sigma$ baryon as a function of $x$ at different values of $s_0$, $M^2=1.3~GeV^2$ and  $T=0$. (b) The same as (a) but for $\Lambda$ hyperon. (c) The same as (a) but for $\Xi$ hyperon and at $M^2=1.5~GeV^2$. } \label{fig1}
\end{figure}

\begin{figure}[h!]
\begin{center}
\subfigure[]{\includegraphics[width=8cm]{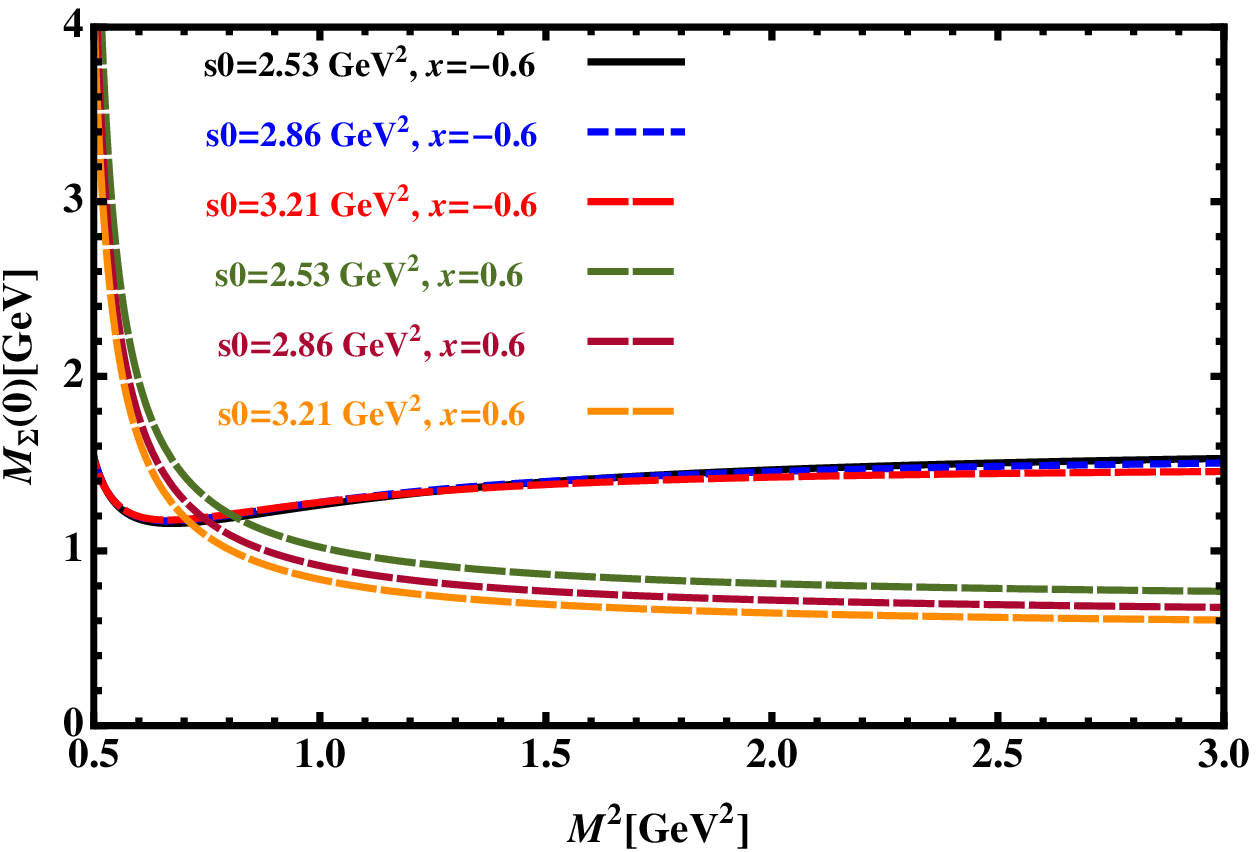}}
\subfigure[]{\includegraphics[width=8cm]{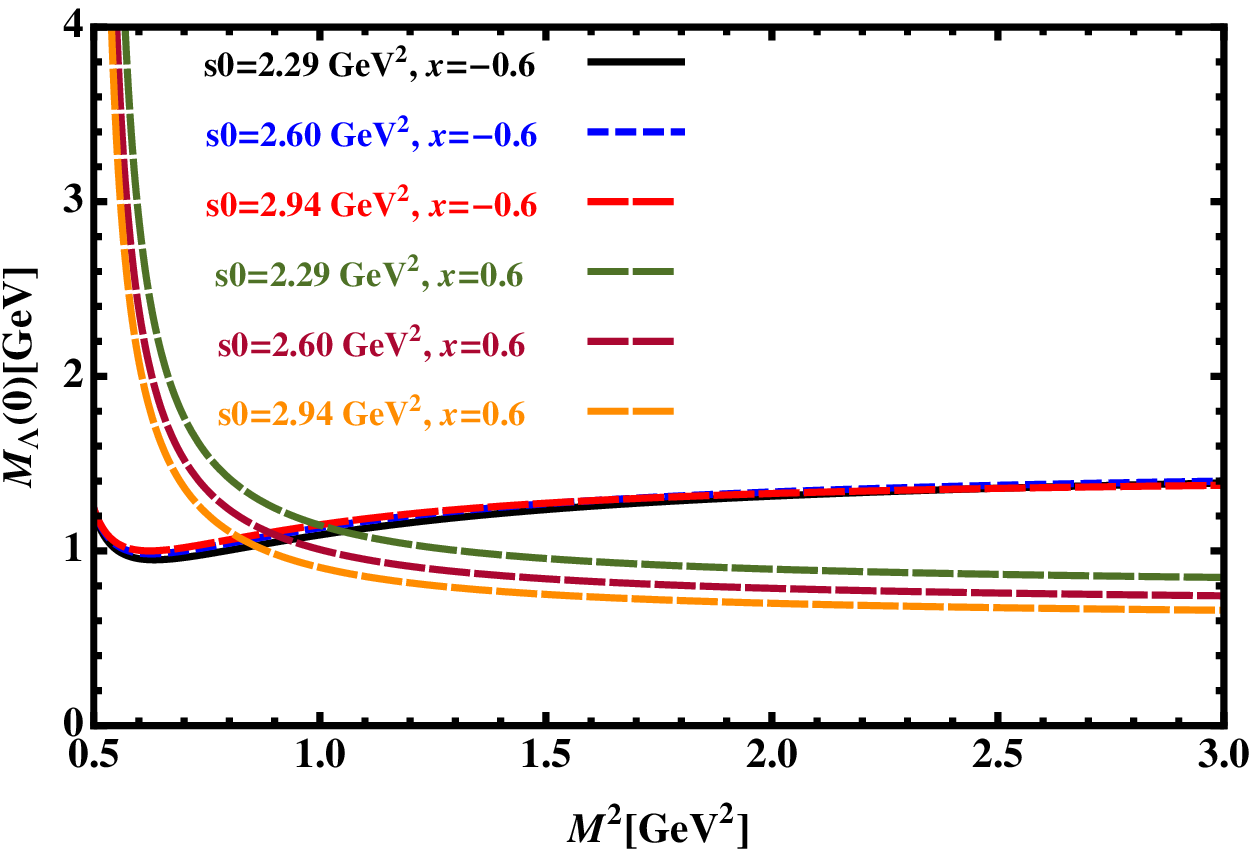}}
\subfigure[]{\includegraphics[width=8cm]{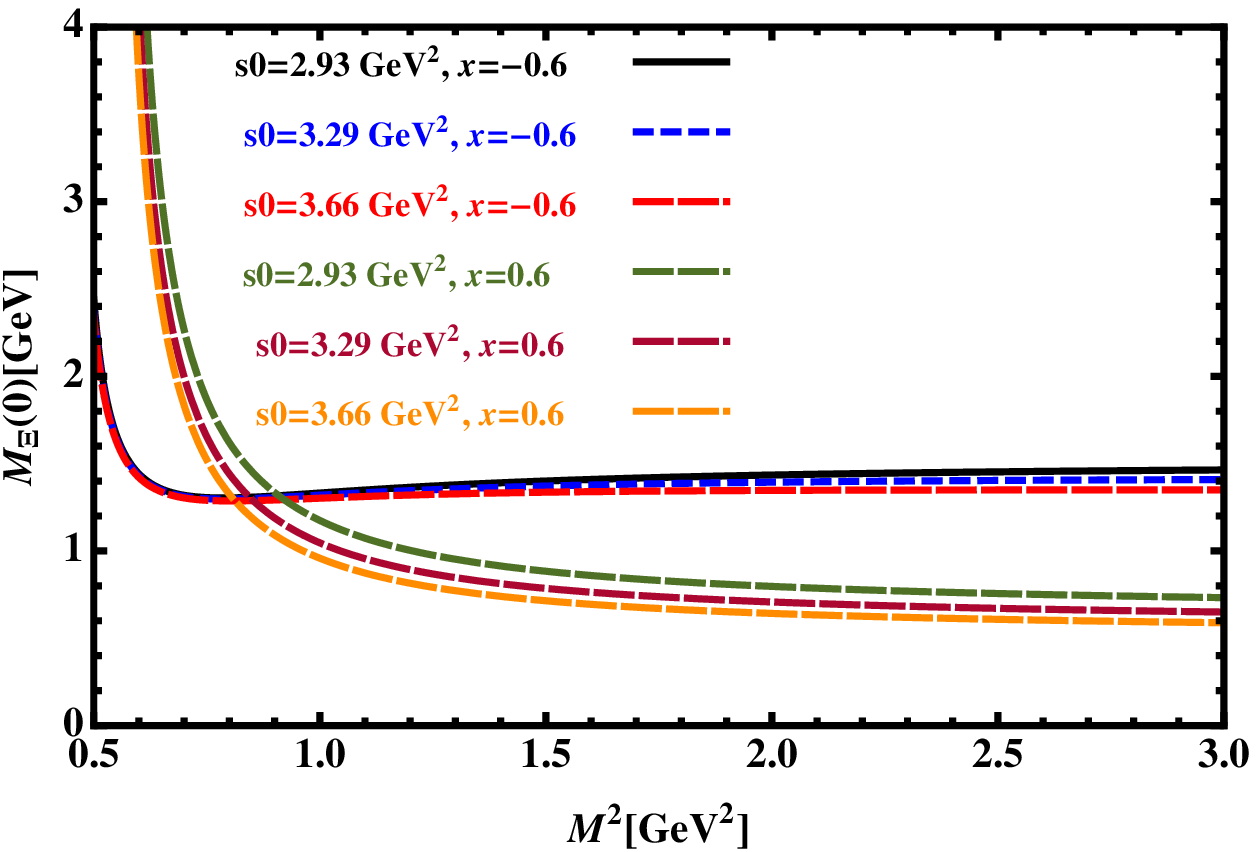}}
\end{center}
\caption{(a) The mass of the  $\Sigma$ baryon as a function of $M^2$ at $T=0$; and at different fixed values of $s_0$ and  $x$. (b) The same as (a) but for $\Lambda$ hyperon.  (c) The same as (a) but for $\Xi$ hyperon.} \label{fig2}
\end{figure}

\begin{figure}[h!]
\begin{center}
\subfigure[]{\includegraphics[width=8cm]{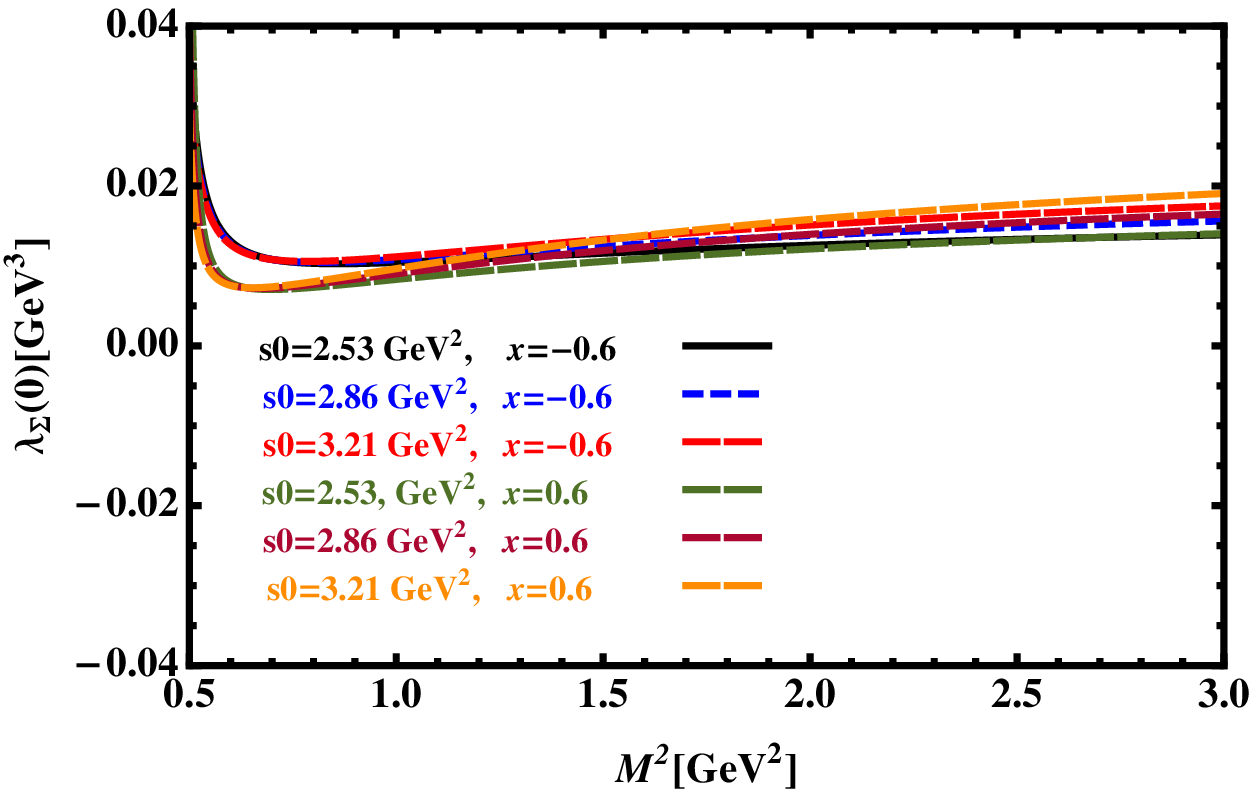}}
\subfigure[]{\includegraphics[width=8cm]{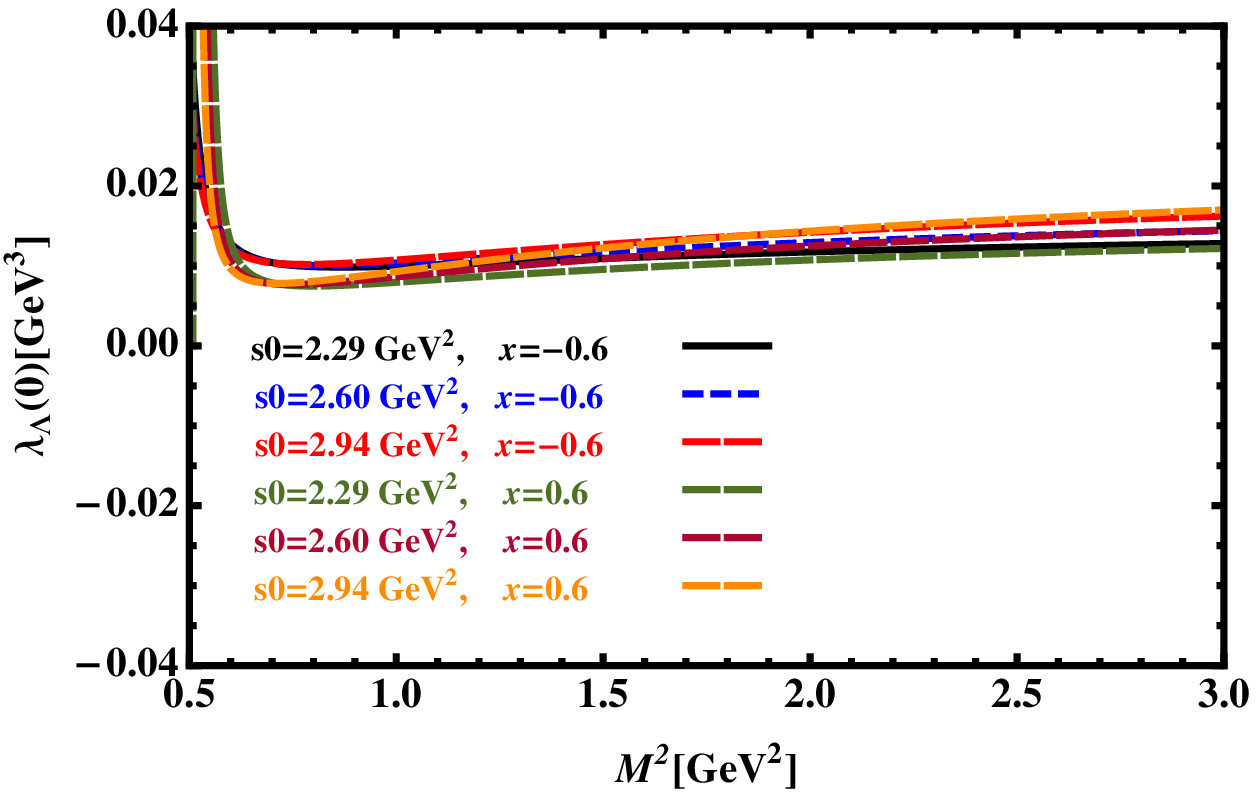}}
\subfigure[]{\includegraphics[width=8cm]{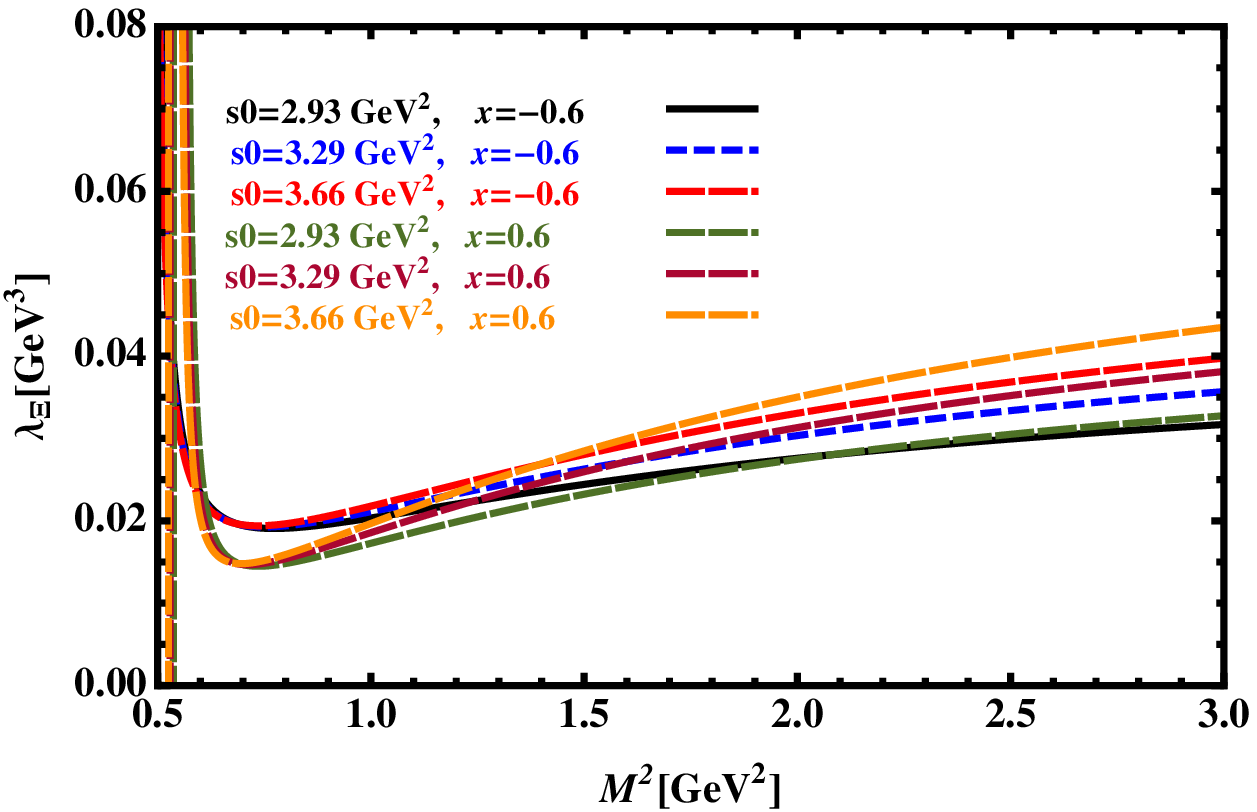}}
\end{center}
\caption{(a) The residue of the  $\Sigma$ baryon as a function of $M^2$ at $T=0$; and at different fixed values of $s_0$ and $x$. (b)  The same as (a) but for $\Lambda$ hyperon. (c)  The same as (a) but for  $\Xi$ hyperon.} \label{fig3}
\end{figure}

By these considerations, we find the working intervals for $M^{2}$ for different members as presented in table \ref{table3}. To determine the working region of the mixing parameter  $t$, we plot the residues
 of the corresponding hyperons with respect to $x=cos\theta$, 
where $t=tan\theta$, at fixed values of the Borel mass parameter $M^{2}$ and $s_0$ at $T=0$ in figure \ref{fig1}. From this figure, we find the working intervals of  $x$ for each baryon presented in table \ref{table3}. 
In these regions, the residues of the corresponding hyperons weakly depend on the parameter $x$ compared to the other regions. From this figure we also see that the results depend on $s_0$ in its working interval very weakly.
 In order to see the dependence of the mass and residue of the hyperons under consideration on the Borel mass 
parameter $M^{2}$, we plot the dependence of $m_{h}(0)$ and $\lambda_{h}(0)$ on $M^{2}$ in a wide region and at different fixed values of the continuum threshold in figures \ref{fig2} and \ref{fig3}.
 From these figures, we see that the mass and residue of 
 $\Sigma$, $\Lambda$ and $\Xi$ hyperons at $T=0$ show good stability with respect to Borel mass parameter $M^{2}$ in its working region for fixed values of $x$. These figures also indicate that the results 
depend on 
choices of $s_0$ very weakly.

\begin{table}[h]
\renewcommand{\arraystretch}{1.5}
\addtolength{\arraycolsep}{3pt}
$$
\begin{array}{|c|c|c|c|}
\hline \hline
         & M^{2}  &x   \\
\hline
  \mbox{$\Sigma$} &1.0~GeV^2\leqslant M^2 \leqslant 1.6~GeV^2&-0.8\leqslant x \leqslant -0.4~\mbox{and}~0.4\leqslant x \leqslant 0.8 \\
  \hline
  \mbox{$\Lambda$} &1.0~GeV^2\leqslant M^2 \leqslant 1.6~GeV^2&-0.8\leqslant x \leqslant -0.4~\mbox{and}~0.4\leqslant x \leqslant 0.8 \\
  \hline
  \mbox{$\Xi$} &1.2~GeV^2\leqslant M^2 \leqslant 1.8~GeV^2&-0.8\leqslant x \leqslant -0.5~\mbox{and}~0.5\leqslant x \leqslant 0.8 \\
                    \hline \hline
\end{array}
$$
\caption{ The working regions of auxiliary parameters $M^{2}$ and $x$ for $\Sigma$, $\Lambda$ and $\Xi$ hyperons.} \label{fitfunction1}
\renewcommand{\arraystretch}{1}
\addtolength{\arraycolsep}{-1.0pt} \label{table3}
\end{table}
 \begin{figure}[h!]
\begin{center}
\subfigure[]{\includegraphics[width=8cm]{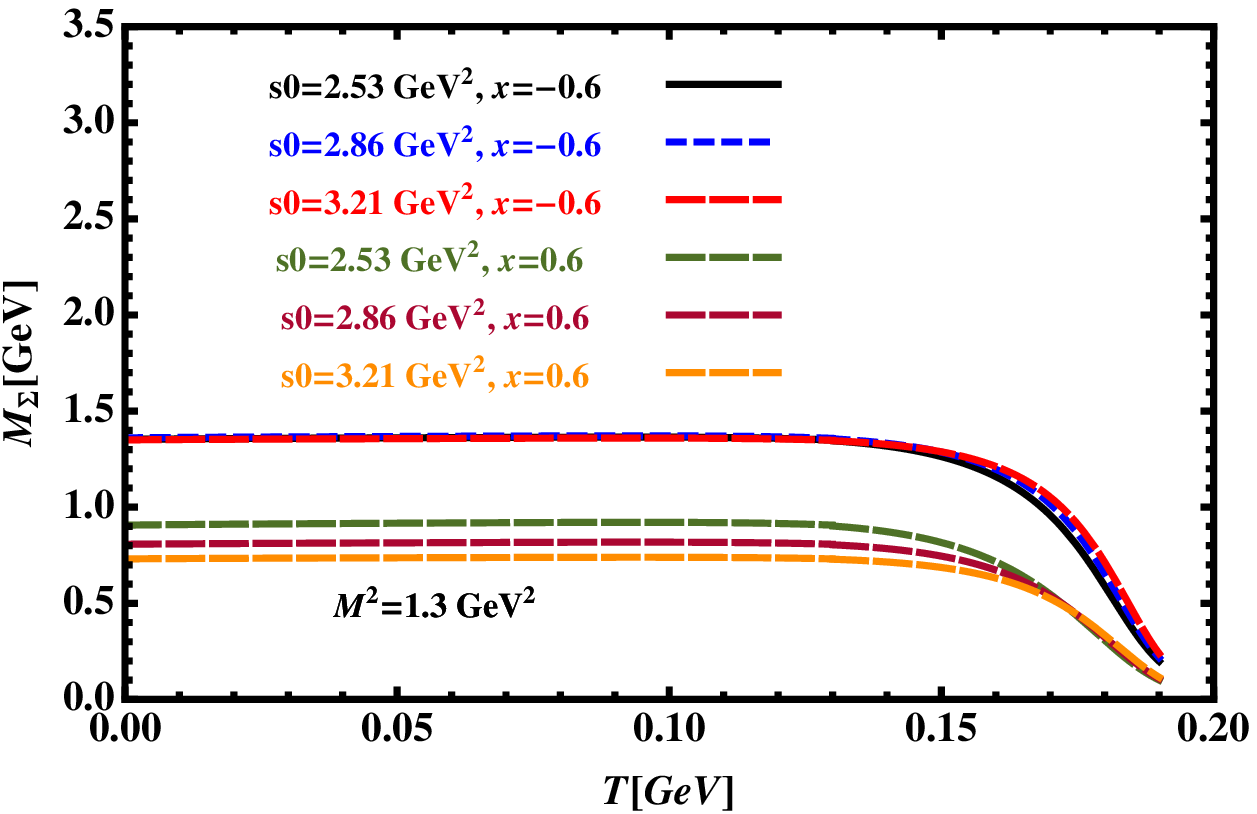}}
\subfigure[]{\includegraphics[width=8cm]{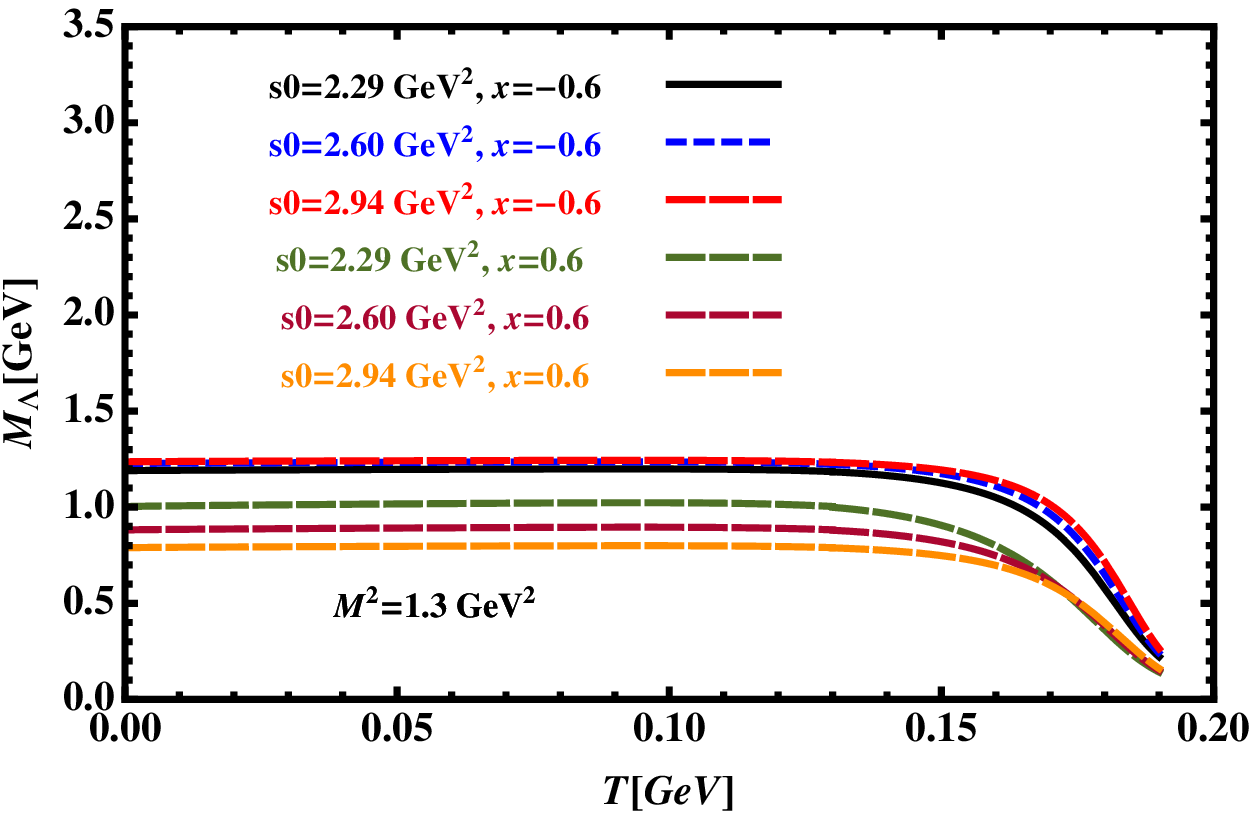}}
\subfigure[]{\includegraphics[width=8cm]{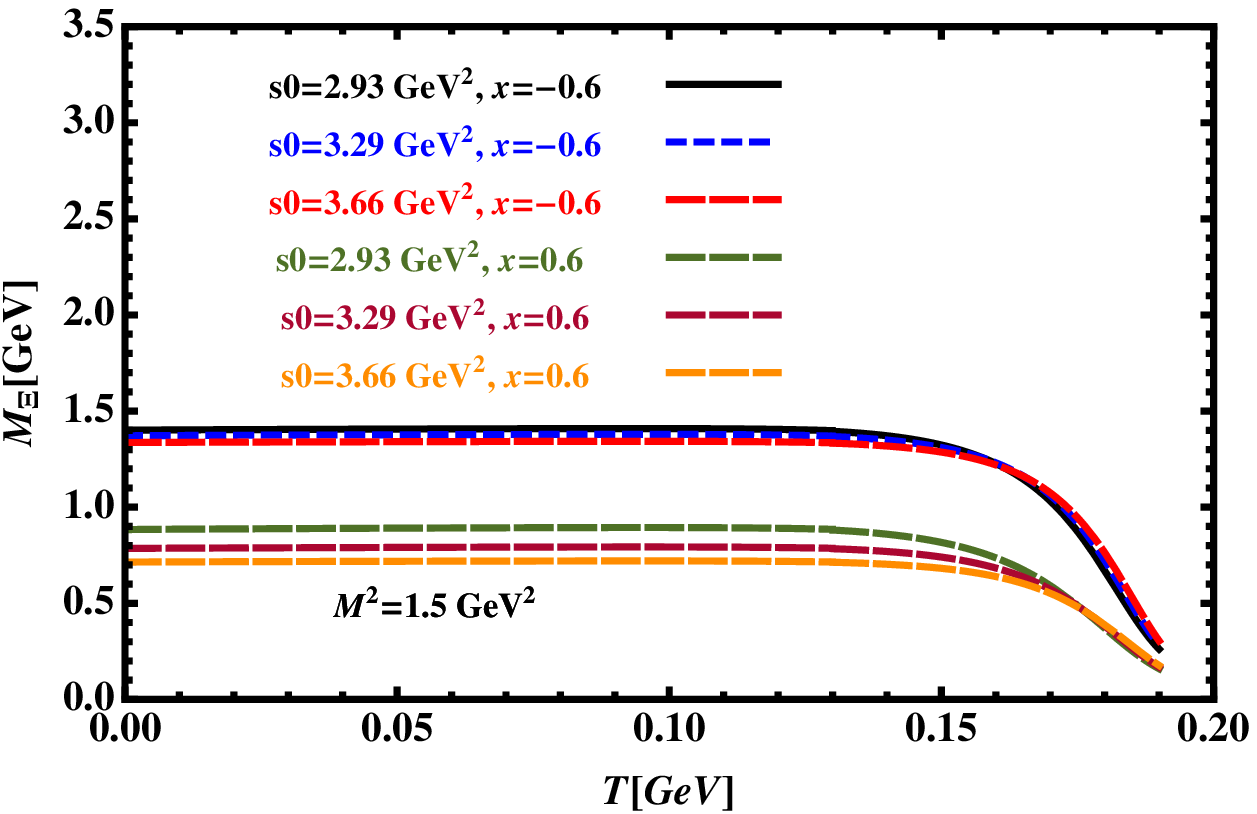}}
\end{center}
\caption{(a) The mass of the  $\Sigma$ baryon as a function of temperature at $M^{2}=1.3~GeV^2$; and at different fixed values of $s_0$ and $x$. (b) The same as (a) but for $\Lambda$ hyperon. (c) The same as (a) but for $\Xi$ hyperon and at $M^{2}=1.5~GeV^2$.} \label{fig4}
\end{figure}
 \begin{figure}[h!]
\begin{center}
\subfigure[]{\includegraphics[width=8cm]{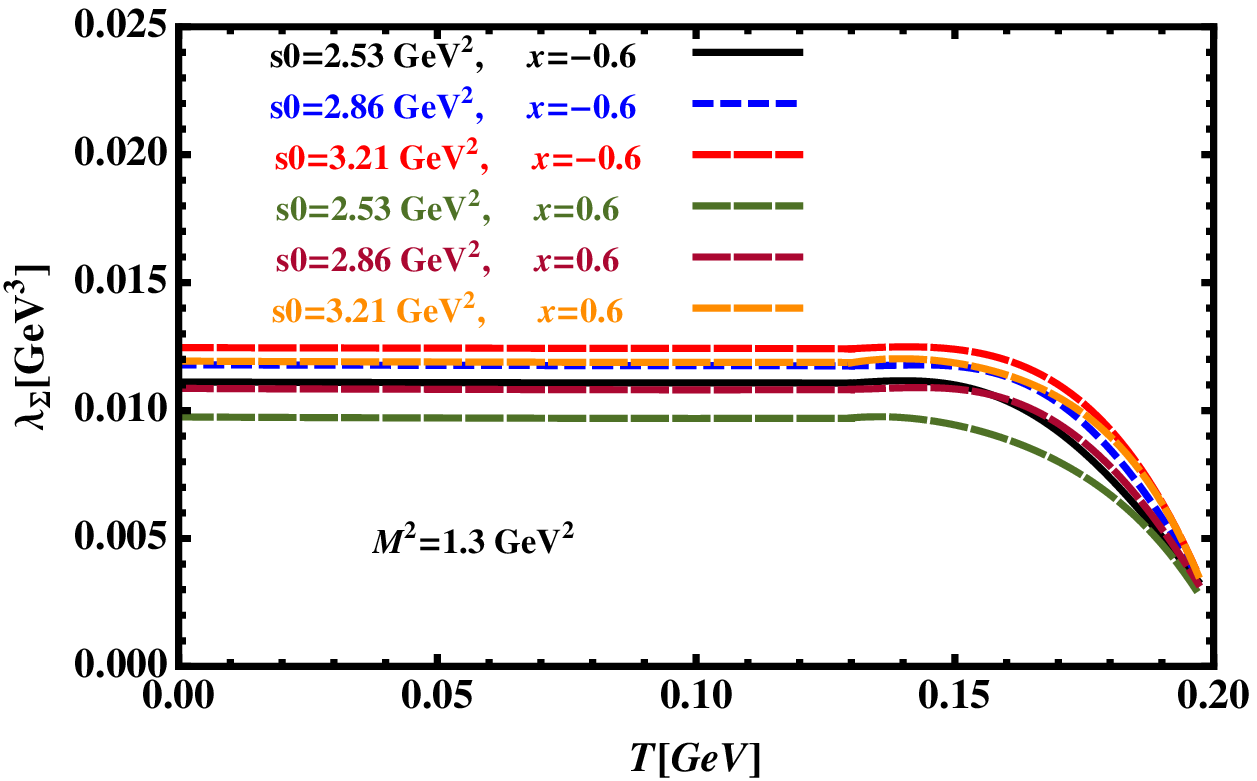}}
\subfigure[]{\includegraphics[width=8cm]{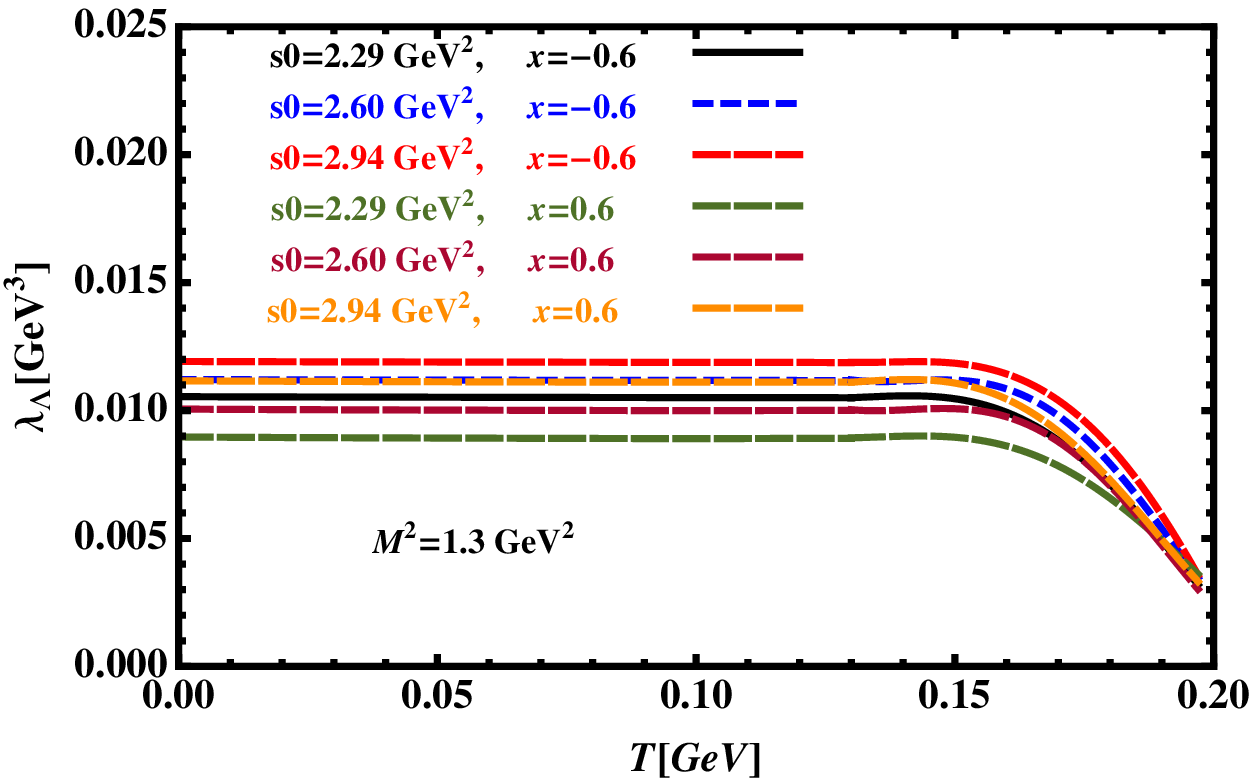}}
\subfigure[]{\includegraphics[width=8cm]{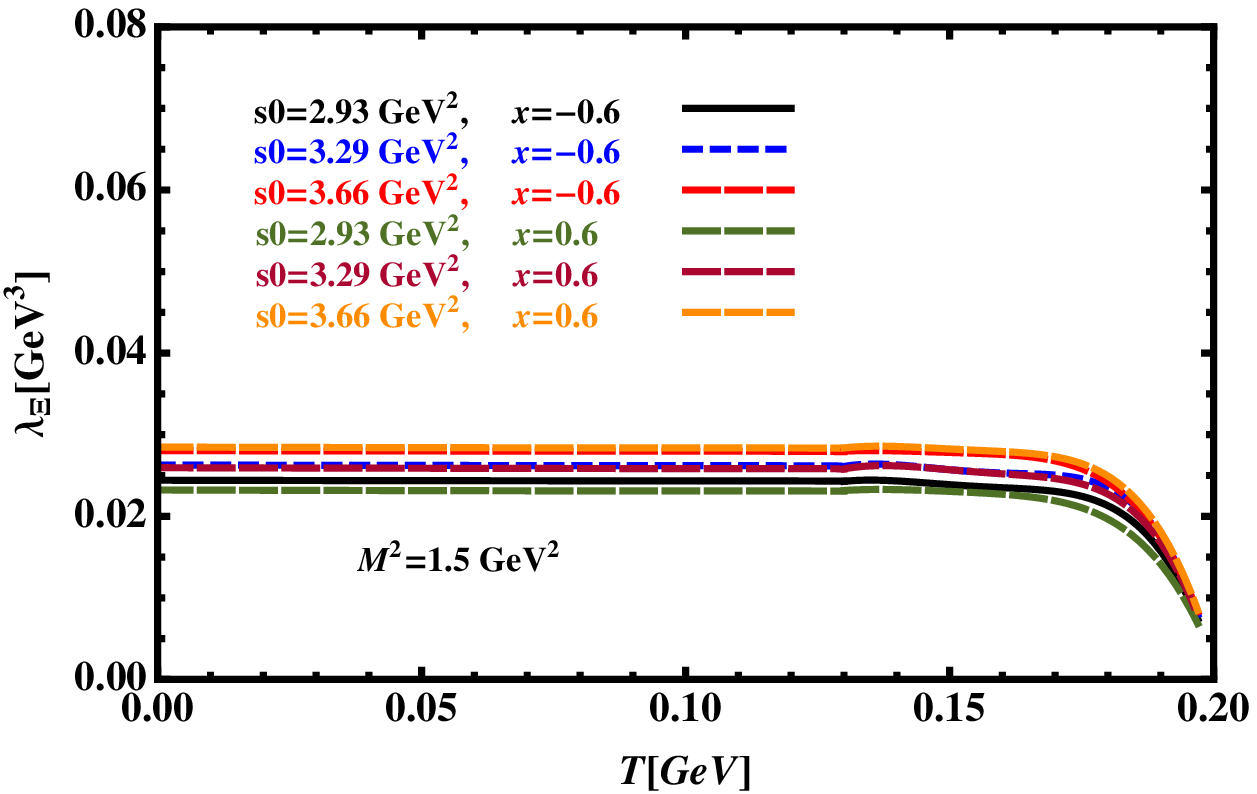}}
\end{center}
\caption{(a) The residue of the  $\Sigma$ baryon as a function of temperature at $M^{2}=1.3~GeV^2$; and at different fixed values of  $s_0$ and $x$. (b) The same as (a) but for $\Lambda$ hyperon. (c) The same as (a) but for $\Xi$ hyperon and at $M^{2}=1.5~GeV^2$.} \label{fig5}
\end{figure}

To investigate the variations of the mass and residue of  $\Sigma$, $\Lambda$ and $\Xi$ hyperons with respect to temperature, we plot these quantities as a function of temperature in 
figures \ref{fig4} and \ref{fig5}. From these figures we observe that the mass and residue of $\Sigma$, $\Lambda$ and $\Xi$ hyperons remain approximately unchanged up to $T\cong0.15~ GeV,$ after which they start
 to rapidly decrease with increasing the temperature. The masses fall to roughly $14\%$, $18\%$ and $21\%$  of their zero-temperature values for  $\Sigma$, $\Lambda$ and $\Xi$ hyperons, respectively
 near to the critical temperature, while the residues overall reach to roughly $30\%$ of their vacuum values near to the critical temperature.
 From these figures, we also read that, at any temperature, the results depend on $s_0$  very weakly in the working interval of this parameter.

Finally, we would like to compare the behaviors of the mass and residue of the hyperons in terms of temperature exhibited in figures \ref{fig4} and \ref{fig5} with our previous work \cite{Azizi} on the finite
 temperature nucleon properties. We observe that  the behaviors of the mass and residue of the hyperons in terms of temperature are similar to those of the nucleon. 
Although, the order of decrease in the values of the masses of the hyperons are about $86\%$, $82\%$ and $79\%$ for   $\Sigma$, $\Lambda$ and $\Xi$ hyperons, respectively   compared to their vacuum values, 
while this order is roughly $58\%$ for the nucleon. In the case of residue,
the residue of hyperons decrease about  $70\%$ compared with their values at $T=0$, while this amount is  $40\%$ for the residue of the nucleon near to the critical temperature. Also, we realize that
 our results for 
the behavior of the mass of hyperons in terms of temperature are in good agreements with those of  \cite{Ryu} for hyperon's mass.


Summarizing, we have calculated the continuum threshold, mass and residue of  $\Sigma$, $\Lambda$ and $\Xi$ hyperons at finite temperature via the thermal QCD sum rules and using the general form of the
 interpolating currents for the corresponding hyperons. We found the reliable working regions of the auxiliary parameters entering the calculations. To numerically analyze the sum rules obtained, 
we used the temperature-dependent quark and gluon condensates as well as the thermal average of the energy density obtained via lattice QCD and QCD sum rules. We observed that the mass and residue of the hyperons decrease near to the 
critical temperature similar to the nucleon's mass and residue previously discussed in \cite{Azizi}, although the order of decreases in the values are different. 
The behavior of the hyperon's mass in terms of temperature in our work is also in agreement with the existing result for the behavior of the mass of the hyperons in \cite{Ryu}. 

The temperature-dependent mass and residue of hyperons obtained in the present work can be used in calculations of many parameters related to the electromagnetic properties
 (charge distribution, magnetic dipole moment, etc.) and  radiative, weak and strong decays of hyperons in hot and dense
medium as well as in determination of the in-medium total width of the considered baryons.
Our results can also be used in analyses of the results of the future heavy ion collision experiments as well as the structure of the dense astrophysical objects like neutron stars as the hyperons, besides the 
nucleons, may be produced
in such experiments and in the core of these stars where the density is high.

\section{Acknowledgment}
This work has been partly supported by the Scientific and Technological
Research Council of Turkey (TUBITAK) under the national postdoctoral research scholarship program 2218.

\newpage

\section*{Appendix A }

In this appendix, we present the functions $\Pi_{p}^{OPE}(p_0,T)$ and  $\Pi_{S}^{OPE}(p_0,T)$ for $\Sigma$, $\Lambda$ and $\Xi$ hyperons in Borel scheme, which are obtained as 

\begin{eqnarray}\label{borelpi1}
\hat{B}\Pi_{p, \Sigma}^{OPE}(p_0,T)&=&-\frac{5 + 2t+ 5t^2}{2048\pi^4}\int^{s_0(T)}_{(m_{u}+m_{d}+m_{s})^2}ds \exp\Big(-\frac{s}{M^2}\Big)s^2\nonumber\\
&+&\frac{\langle\bar{q}q\rangle}{384\pi^2}\Bigg[2m_{0}^2\Big((1 + 4t+ t^2)m_{d} - 9(-1 + t^2)m_{s}+(1 + 4t + t^2)m_{u}\Big)\nonumber\\
&-&(1 + t)\Big(m_{d}+ 4m_{s}+ m_{u} + t (m_{d}  -4 m_{s} +  m_{u})\Big)\nonumber\\
&\times&\int^{s_0(T)}_{(m_{u}+m_{d}+m_{s})^2}ds\exp\Big(-\frac{s}{M^2}\Big)\Bigg]\nonumber\\
&+&\frac{\langle\bar{s}s\rangle}{384\pi^2}\Bigg[m_{0}^2\Big(-9(-1 + t^2)m_{d} +(5 + 2t + 5t^2)m_{s}-9(-1 + t^2)m_{u}\Big)\nonumber\\
&+&3(6(-1 + t^2)m_{d} - (5 + 2t + 5t^2)m_{s}+ 6(-1 + t^2) m_{u})\nonumber\\
&\times&\int^{s_0(T)}_{(m_{u}+m_{d}+m_{s})^2}ds\exp\Big(-\frac{s}{M^2}\Big)\Bigg]\nonumber\\
&-&\frac{\langle u\Theta^{f} u \rangle}{288\pi^2}(5 + 2t+ 5t^2)\Bigg[8p_{0}^2-5\int^{s_0(T)}_{(m_{u}+m_{d}+m_{s})^2}ds\exp\Big(-\frac{s}{M^2}\Big)\Bigg]\nonumber\\
&+&\frac{\alpha_{s} \langle u\Theta^{g} u \rangle}{768\pi^3 N_{c}}(5 + 2t + 5t^2)(-1 +N_{c})\Bigg[4p_{0}^2 -\int^{s_0(T)}_{(m_{u}+m_{d}+m_{s})^2}ds\exp\Big(-\frac{s}{M^2}\Big)\Bigg]\nonumber\\
&+&\frac{3\langle\alpha_{s}G^2\rangle}{1024\pi^3 N_{c}}(5 + 2t+ 5t^2)(-1 + N_{c})\int^{s_0(T)}_{(m_{u}+m_{d}+m_{s})^2}ds\exp\Big(-\frac{s}{M^2}\Big)\nonumber\\
&+&\langle\bar{q}q\rangle^2\Bigg[\frac{(-1 + t)^2 (m_{0}^2 - 2M^2)}{48M^2}\Bigg]\nonumber \\
&+&\langle\bar{q}q\rangle \langle\bar{s}s\rangle \Bigg[\frac{(-1 + t^2) (m_{0}^2 - 2M^2)}{8M^2}\Bigg]\nonumber \\
&+&\frac{\langle\bar{q}q\rangle\langle u\Theta^{f} u \rangle}{54M^6}\Bigg\{(2 + t + 2t^2)m_{0}^2 (m_{d}+ m_{u})(M^2 + 2p_{0}^2)  \nonumber\\
&-&3M^2 \Bigg[  3(-1 + t^2)m_{s}M^2+2 m_{d}  \Big( (1 + t^2)M^2 +2(2 + t + 2t^2)p_{0}^2\Big)\nonumber\\
&+&2m_{u} \Big((1 + t^2)M^2 + 2(2 + t + 2t^2)p_{0}^2\Big)\Bigg]\Bigg\}\nonumber\\
&+&\frac{\langle\bar{s}s\rangle\langle u\Theta^{f} u \rangle}{108M^6}\Bigg\{2(1 + t^2)m_{0}^2 m_{s}(M^2 + 2p_{0}^2) \nonumber\\
&+&3M^2\Bigg[-3(-1 + t^2)m_{d} M^2-3(-1 + t^2)m_{u}M^2 \nonumber\\
&+& m_{s}\Big((1 + t)^2M^2- 8(1 + t^2)p_{0}^2\Big)\Bigg]\Bigg\}\nonumber\\
&+&\frac{\alpha_{s}\langle\bar{q}q\rangle\langle u\Theta^{g} u \rangle}{576\pi N_{c}M^6}\Bigg[(3 + 2t + 3t^2)(m_{d}+ m_{u})(-1 + N_{c})\nonumber\\
&\times&\Big(m_{0}^2(3M^2-2p_{0}^2)+ 3M^2(-5M^2+4p_{0}^2)\Big)\Bigg]\nonumber\\
&+&\frac{\alpha_{s}\langle\bar{s}s\rangle\langle u\Theta^{g} u \rangle}{576\pi N_{c}M^6}\Bigg[(1 +t)^2 m_{s}(-1 + N_{c})\nonumber\\
&\times&\Big(3M^2(5M^2 - 4p_{0}^2) + m_{0}^2(-3M^2+2p_{0}^2)\Big)\Bigg]\nonumber\\
&+&\frac{\langle\bar{q}q\rangle\langle\alpha_{s}G^2\rangle}{768\pi N_{c}M^4}(3 + 2t + 3t^2)(m_{d}+ m_{u})(4m_{0}^2 - 15M^2)(-1 + N_{c}) \nonumber\\
&-&\frac{\langle\bar{s}s\rangle\langle\alpha_{s}G^2\rangle}{768\pi N_{c}M^4}(1 + t)^2 m_{s}(4m_{0}^2 - 15M^2)(-1 + N_{c}) \nonumber\\
&-&\frac{\alpha_{s} \langle u\Theta^{f} u \rangle \langle u\Theta^{g} u \rangle}{144\pi N_{c}M^4}(-1 + N_{c})\Big[(21 + 2t+ 21t^2)M^2 - 4(7 + 6t+ 7t^2)p_{0}^2\Big]\nonumber\\
&-&\frac{\langle\alpha_{s}G^2\rangle \langle u\Theta^{f} u \rangle}{576\pi N_{c}M^4}(-1 + N_{c})\Big[(71 + 22t + 71t^2)M^2 - 8(1 + t)^2p_{0}^2\Big] \nonumber\\
&-&\frac{2\langle u\Theta^{f} u \rangle^2}{18 M^2}(5 + 2t + 5t^2), \end{eqnarray}
\begin{eqnarray}\label{borelpi2}
\hat{B}\Pi_{p, \Lambda}^{OPE}(p_0,T)&=&\frac{5 + 2t+ 5t^2}{2048\pi^4}\int^{s_0(T)}_{(m_{u}+m_{d}+m_{s})^2}ds \exp\Big(-\frac{s}{M^2}\Big)s^2\nonumber\\
&+&\frac{\langle\bar{q}q\rangle}{192\pi^2}(-1 - 4t+ 5t^2)m_{s}\Bigg[m_{0}^2-2\int^{s_0(T)}_{(m_{u}+m_{d}+m_{s})^2}ds\exp\Big(-\frac{s}{M^2}\Big)\Bigg]\nonumber\\
&-&\frac{\langle\bar{s}s\rangle}{384\pi^2}(5 + 2t + 5t^2)m_{s}\Bigg[m_{0}^2-3\int^{s_0(T)}_{(m_{u}+m_{d}+m_{s})^2}ds\exp\Big(-\frac{s}{M^2}\Big)\Bigg]\nonumber\\
&+&\frac{\langle u\Theta^{f} u \rangle}{288\pi^2}(5 + 2t+ 5t^2)\Bigg[8p_{0}^2-5\int^{s_0(T)}_{(m_{u}+m_{d}+m_{s})^2}ds\exp\Big(-\frac{s}{M^2}\Big)\Bigg]\nonumber\\
&-&\frac{\alpha_{s} \langle u\Theta^{g} u \rangle}{768\pi^3 N_{c}}(5 + 2t + 5t^2)(-1 +N_{c})\Bigg[4p_{0}^2 -\int^{s_0(T)}_{(m_{u}+m_{d}+m_{s})^2}ds\exp\Big(-\frac{s}{M^2}\Big)\Bigg]\nonumber\\
&-&\frac{3\langle\alpha_{s}G^2\rangle}{1024\pi^3 N_{c}}(5 + 2t+ 5t^2)(-1 + N_{c})\int^{s_0(T)}_{(m_{u}+m_{d}+m_{s})^2}ds\exp\Big(-\frac{s}{M^2}\Big)\nonumber\\
&-&\langle\bar{q}q\rangle^2\Bigg[\frac{(-13 + 2t + 11t^2)(m_{0}^2 - 2M^2)}{144M^2}\Bigg]\nonumber \\
&-&\langle\bar{q}q\rangle \langle\bar{s}s\rangle \Bigg[\frac{(-1 - 4t+ 5t^2)(m_{0}^2 - 2M^2)}{72M^2}\Bigg]\nonumber \\
&+&\frac{\langle\bar{q}q\rangle\langle u\Theta^{f} u \rangle}{54M^2}(-1 - 4t + 5t^2)m_{s})\nonumber\\
&+&\frac{\langle\bar{s}s\rangle\langle u\Theta^{f} u \rangle}{324M^6}\Bigg\{m_{s}\Bigg[-2(7 + 4t + 7t^2)m_{0}^2 (M^2 + 2p_{0}^2) \nonumber\\
&+&3M^2\Big((13 + 10t + 13t^2) M^2 + 8(7 + 4t + 7t^2)p_{0}^2\Big)\Bigg]\Bigg\}\nonumber\\
&+&\frac{\alpha_{s}\langle\bar{s}s\rangle\langle u\Theta^{g} u \rangle}{1728\pi N_{c}M^6}\Bigg[(13 + 10t + 13t^2) m_{s}(-1 + N_{c})\nonumber\\
&\times&\Big(3M^2(5M^2 - 4p_{0}^2) + m_{0}^2(-3M^2+2p_{0}^2)\Big)\Bigg]\nonumber\\
&-&\frac{\langle\bar{s}s\rangle\langle\alpha_{s}G^2\rangle}{2304\pi N_{c}M^4}(13 + 10t + 13t^2) m_{s}(4m_{0}^2 - 15M^2)(-1 + N_{c}) \nonumber\\
&-&\frac{\alpha_{s} \langle u\Theta^{f} u \rangle \langle u\Theta^{g} u \rangle}{144\pi N_{c}M^4}(-1 + N_{c})\Big[(21 + 2t+ 21t^2)M^2 - 4(7 + 6t+ 7t^2)p_{0}^2\Big]\nonumber\\
&-&\frac{\langle\alpha_{s}G^2\rangle \langle u\Theta^{f} u \rangle}{576\pi N_{c}M^4}(-1 + N_{c})\Big[(71 + 22t + 71t^2)M^2 - 8(1 + t)^2p_{0}^2\Big]\nonumber\\
&+&\frac{\langle u\Theta^{f} u \rangle^2}{18 M^2}(5 + 2t + 5t^2),
\end{eqnarray}
\begin{eqnarray}\label{borelpi3}
\hat{B}\Pi_{p, \Xi}^{OPE}(p_0,T)&=&-\frac{5 + 2t+ 5t^2}{512\pi^4}\int^{s_0(T)}_{(2m_{s}+m_{u})^2}ds \exp\Big(-\frac{s}{M^2}\Big)s^2\nonumber\\
&+&\frac{\langle\bar{q}q\rangle}{96\pi^2}\Bigg[m_{0}^2\Big(-18(-1 + t^2)m_{s}+(5 + 2t +5 t^2)m_{u}\Big)\nonumber\\
&+&3\Big(12(-1 + t^2)m_{s} - (5 + 2t+ 5t^2) m_{u})\Big)\int^{s_0(T)}_{(2m_{s}+m_{u})^2}ds\exp\Big(-\frac{s}{M^2}\Big)\Bigg]\nonumber\\
&+&\frac{\langle\bar{s}s\rangle}{48\pi^2}\Bigg[m_{0}^2\Big(2(1 + 4t + t^2)m_{s}-9(-1 + t^2)m_{u}\Big)\nonumber\\
&-&9(1+t)\Big(m_{s}(1+t)+ 2m_{u}(1-t)\Big)\int^{s_0(T)}_{(2m_{s}+m_{u})^2}ds\exp\Big(-\frac{s}{M^2}\Big)\Bigg]\nonumber\\
&-&\frac{\langle u\Theta^{f} u \rangle}{72\pi^2}(5 + 2t+ 5t^2)\Bigg[8p_{0}^2-5\int^{s_0(T)}_{(2m_{s}+m_{u})^2}ds\exp\Big(-\frac{s}{M^2}\Big)\Bigg]\nonumber\\
&+&\frac{\alpha_{s} \langle u\Theta^{g} u \rangle}{192\pi^3 N_{c}}(5 + 2t + 5t^2)(-1 +N_{c})\Bigg[4p_{0}^2 -\int^{s_0(T)}_{(2m_{s}+m_{u})^2}ds\exp\Big(-\frac{s}{M^2}\Big)\Bigg]\nonumber\\
&+&\frac{3\langle\alpha_{s}G^2\rangle}{256\pi^3 N_{c}}(5 + 2t+ 5t^2)(-1 + N_{c})\int^{s_0(T)}_{(2m_{s}+m_{u})^2}ds\exp\Big(-\frac{s}{M^2}\Big)\nonumber\\
&+&\langle\bar{s}s\rangle^2 \Bigg[\frac{(-1 + t)^2 (m_{0}^2 - 2M^2)}{12M^2}\Bigg]\nonumber \\
&+&\langle\bar{q}q\rangle \langle\bar{s}s\rangle \Bigg[\frac{(-1 + t^2) (m_{0}^2 - 2M^2)}{2M^2}\Bigg]\nonumber \\
&+&\frac{\langle\bar{q}q\rangle\langle u\Theta^{f} u \rangle}{27M^6}\Bigg\{2(1 + t^2) m_{0}^2  m_{u}(M^2 + 2p_{0}^2)  \nonumber\\
&+&3M^2\Bigg[ -6(-1 + t^2)m_{s}M^2+m_{u}\Big((1 + t)^2M^2 -8(1 + t^2)p_{0}^2\Big)\Bigg]\Bigg\}\nonumber\\
&+&2\frac{\langle\bar{s}s\rangle\langle u\Theta^{f} u \rangle}{27M^6}\Bigg\{2(2 + t+2t^2)m_{0}^2 m_{s}(M^2 + 2p_{0}^2) \nonumber\\
&-&3M^2\Bigg[3(-1 + t^2)m_{u}M^2 + 4m_{s}\Big((1 + t^2)M^2+2(2 +t+ 2t^2)p_{0}^2\Big)\Bigg]\Bigg\}\nonumber\\
&+&\frac{\alpha_{s}\langle\bar{q}q\rangle\langle u\Theta^{g} u \rangle}{144\pi N_{c}M^6}\Bigg[(1 + t^2) m_{u}(-1 + N_{c})\nonumber\\
&\times&\Big(3M^2(5M^2-4p_{0}^2)+m_{0}^2(-3M^2+2p_{0}^2)\Big)\Bigg]\nonumber\\
&+&\frac{\alpha_{s}\langle\bar{s}s\rangle\langle u\Theta^{g} u \rangle}{72\pi N_{c}M^6}\Bigg[(3 + 2t + 3t^2) m_{s}(-1 + N_{c})\nonumber\\
&\times&\Big(m_{0}^2(3M^2-2p_{0}^2) +3M^2(-5M^2 + 4p_{0}^2)\Big)\Bigg]\nonumber\\
&-&\frac{\langle\bar{q}q\rangle\langle\alpha_{s}G^2\rangle}{192\pi N_{c}M^4}(1 + t)^2 m_{u}(4m_{0}^2 - 15M^2)(-1 + N_{c}) \nonumber\\
&+&\frac{\langle\bar{s}s\rangle\langle\alpha_{s}G^2\rangle}{96\pi N_{c}M^4}(3 + 2t + 3t^2) m_{s}(4m_{0}^2 - 15M^2)(-1 + N_{c}) \nonumber\\
&+&\frac{\alpha_{s} \langle u\Theta^{f} u \rangle \langle u\Theta^{g} u \rangle}{36\pi N_{c}M^4}(-1 + N_{c})\Big[(21 + 2t+ 21t^2)M^2 - 4(7 + 6t+ 7t^2)p_{0}^2\Big]\nonumber\\
&+&\frac{\langle\alpha_{s}G^2\rangle \langle u\Theta^{f} u \rangle}{144\pi N_{c}M^4}(-1 + N_{c})\Big[(71 + 22t + 71t^2)M^2 - 8(1 + t)^2p_{0}^2\Big]\nonumber\\
&-&\frac{2\langle u\Theta^{f} u \rangle^2}{9 M^2}(5 + 2t + 5t^2),
\end{eqnarray}
\begin{eqnarray}\label{borelpi1U}
\hat{B}\Pi_{S, \Sigma}^{OPE}(p_0,T)&=&-\frac{(-1 + t)}{512\pi^4}\Bigg[3(1 + t)m_{d} + (-1 + t)m_{s}+3(1 + t)m_{u}\Bigg] \nonumber\\
&\times&  \int^{s_0(T)}_{(m_{u}+m_{d}+m_{s})^2}ds \exp\Big(-\frac{s}{M^2}\Big)s^2\nonumber\\
&-&\frac{3\langle\bar{q}q\rangle}{64\pi^2}(-1 + t^2)\Bigg[ \int^{s_0(T)}_{(m_{u}+m_{d}+m_{s})^2}(m_{0}^2-2s) ds\exp\Big(-\frac{s}{M^2}\Big)\Bigg]\nonumber\\
&-&\frac{\langle\bar{s}s\rangle}{128\pi^2}(-1 + t)^2\Bigg[ \int^{s_0(T)}_{(m_{u}+m_{d}+m_{s})^2}(m_{0}^2-2s) ds\exp\Big(-\frac{s}{M^2}\Big)\Bigg]\nonumber\\
&-&\frac{\langle u\Theta^{f} u \rangle}{6\pi^2}(-1+ t^2)(m_{d}+m_{u})p_{0}^2\nonumber\\
&+&\frac{\alpha_{s} \langle u\Theta^{g} u \rangle}{128\pi^3 N_{c}}(-1 + t)\Bigg[5(1 +t)m_{d}+m_{s} - t m_{s} + 5(1 + t)m_{u}\Bigg](-1 +N_{c})p_{0}^2 \nonumber\\
&+&\frac{3\langle\alpha_{s}G^2\rangle}{512\pi^3 N_{c}}(-1 + t)\Bigg[3(1 +t)m_{d}+m_{s} - t m_{s} + 3(1 + t)m_{u}\Bigg] \nonumber\\
&\times&(-1 +N_{c}) \int^{s_0(T)}_{(m_{u}+m_{d}+m_{s})^2}ds\exp\Big(-\frac{s}{M^2}\Big)\nonumber\\
&+&\langle\bar{q}q\rangle^2\Bigg[\frac{3(-1 + t^2)m_{d}- 2(5 + 2t + 5t^2)m_{s} + 3(-1 + t^2)m_{u}}{48}\Bigg]\nonumber \\
&-&\langle\bar{q}q\rangle \langle\bar{s}s\rangle \Bigg[\frac{(3 + 2t + 3t^2)m_{d} - 2(-1 + t^2)m_{s} + (3 + 2t + 3t^2)m_{u}}{16}\Bigg]\nonumber \\
&+&\frac{\langle\bar{q}q\rangle\langle u\Theta^{f} u \rangle}{18M^4}(-1 + t^2)(-3m_{0}^2 M^2 + 6M^4 - 4m_{0}^2 p_{0}^2 + 16M^2 p_{0}^2)\nonumber\\
&-&\frac{\langle\bar{s}s\rangle\langle u\Theta^{f} u \rangle}{36M^2}(-1 + t)^2 (m_{0}^2-2M^2)\nonumber\\
&+&\frac{\alpha_{s}\langle\bar{q}q\rangle\langle u\Theta^{g} u \rangle}{96\pi N_{c}M^4}(-1 + t^2)(-1 +N_{c})(-6m_{0}^2 M^2 + 18 M^4 + 5m_{0}^2 p_{0}^2 -20M^2 p_{0}^2)\nonumber\\
&-&\frac{\alpha_{s}\langle\bar{s}s\rangle\langle u\Theta^{g} u \rangle}{192\pi N_{c}M^4}(-1 + t)^2(-1 +N_{c})(-2m_{0}^2 M^2 + 6 M^4 +m_{0}^2 p_{0}^2 -4M^2 p_{0}^2)\nonumber\\
&-&\frac{9\langle\bar{q}q\rangle\langle\alpha_{s}G^2\rangle}{128\pi N_{c}M^2}(-1 + t^2) (m_{0}^2-2M^2) (-1 +N_{c})\nonumber\\
&-&\frac{3\langle\bar{s}s\rangle\langle\alpha_{s}G^2\rangle}{256\pi N_{c}M^2}(-1 + t)^2(m_{0}^2-2M^2)(-1 +N_{c}) \nonumber\\
&-&\frac{2\langle u\Theta^{f} u \rangle^2}{27 M^4}(-1 + t)\Bigg[(-1 + t) m_{s}(M^2 + 4p_{0}^2) \nonumber\\
&+& (1 + t)(3M^2+ 8p_{0}^2) (m_{d}+m_{u})\Bigg],
\end{eqnarray}
\begin{eqnarray}\label{borelpi2U}
\hat{B}\Pi_{S, \Lambda}^{OPE}(p_0,T)&=&\frac{(-13 + 2t + 11t^2)m_{s}}{1536\pi^4}\int^{s_0(T)}_{(m_{u}+m_{d}+m_{s})^2}ds \exp\Big(-\frac{s}{M^2}\Big)s^2\nonumber\\
&+&\frac{\langle\bar{q}q\rangle}{192\pi^2}(-1 - 4t + 5t^2)\Bigg[ \int^{s_0(T)}_{(m_{u}+m_{d}+m_{s})^2}(m_{0}^2-2s) ds\exp\Big(-\frac{s}{M^2}\Big)\Bigg]\nonumber\\
&+&\frac{\langle\bar{s}s\rangle}{384\pi^2}(-13 + 2t + 11t^2)\Bigg[ \int^{s_0(T)}_{(m_{u}+m_{d}+m_{s})^2}(m_{0}^2-2s)ds\exp\Big(-\frac{s}{M^2}\Big)\Bigg]\nonumber\\
&+&\frac{\langle u\Theta^{f} u \rangle}{9\pi^2}m_{s}(-2 +2 t^2)p_{0}^2\nonumber\\
&-&\frac{\alpha_{s} \langle u\Theta^{g} u \rangle}{384\pi^3 N_{c}}(-19 - 2t + 21t^2)m_{s}(-1 +N_{c})p_{0}^2 \nonumber\\
&-&\frac{\langle\alpha_{s}G^2\rangle}{512\pi^3 N_{c}}(-11 - 2t + 13t^2)(-1 +N_{c})m_{s} \int^{s_0(T)}_{(m_{u}+m_{d}+m_{s})^2}ds\exp\Big(-\frac{s}{M^2}\Big)\nonumber\\
&+&\langle\bar{q}q\rangle^2\Bigg[\frac{(5 + 2t + 5t^2)m_{s}}{24}\Bigg]\nonumber \\
&+&\langle\bar{q}q\rangle \langle\bar{s}s\rangle \Bigg[\frac{(1 + 4t - 5t^2)m_{s}}{72}\Bigg]\nonumber \\
&+&\frac{\langle\bar{q}q\rangle\langle u\Theta^{f} u \rangle}{54M^4}(-1 + t)\Bigg[m_{0}^2 ( M^2 + 5t M^2 + 4(1 +t) p_{0}^2)\nonumber\\
&-&2M^2(M^2 + 5tM^2 + 8(1 + t)  p_{0}^2)\Bigg]\nonumber\\
&-&\frac{\langle\bar{s}s\rangle\langle u\Theta^{f} u \rangle}{108M^4}(-1 + t) \Bigg[m_{0}^2 \Big((13 + 11t)M^2 + 16(1 + t)p_{0}^2\Big)\nonumber\\
&-&2M^2\Big((13 + 11t)M^2 + 32(1 + t)p_{0}^2\Big)\Bigg]\nonumber\\
&+&\frac{\alpha_{s}\langle\bar{q}q\rangle\langle u\Theta^{g} u \rangle}{288\pi N_{c}M^4}(-1 + t)(-1 +N_{c}) \Bigg[m_{0}^2\Big(2(5 +t)M^2 - (7 + 3t) p_{0}^2\Big)\nonumber\\
&+&2M^2\Big(-3(5 + t)M^2 + 2(7 + 3t) p_{0}^2\Big)\Bigg]\nonumber\\
&+&\frac{\alpha_{s}\langle\bar{s}s\rangle\langle u\Theta^{g} u \rangle}{192\pi N_{c}M^4}(-1 + t)(-1 +N_{c}) \Bigg[m_{0}^2 \Big((22 + 26t)M^2 - (19 +21t)p_{0}^2\Big)\nonumber\\
&+&2M^2\Big(-3(11 + 13t)M^2 + 2(19 + 21t)p_{0}^2\Big)\Bigg]\nonumber\\
&+&\frac{\langle\bar{q}q\rangle\langle\alpha_{s}G^2\rangle}{128\pi N_{c}M^2}(-5 + 4t +t^2) (m_{0}^2-2M^2) (-1 +N_{c})\nonumber\\
&+&\frac{\langle\bar{s}s\rangle\langle\alpha_{s}G^2\rangle}{256\pi N_{c}M^2}(-11 - 2t + 13t^2)(m_{0}^2-2M^2)(-1 +N_{c}) \nonumber\\
&+&\frac{2\langle u\Theta^{f} u \rangle^2}{81 M^4}(-1 + t)m_{s}\Bigg[(13 + 11t)M^2 + 4(9 + 7t) p_{0}^2\Bigg],
\end{eqnarray}
\begin{eqnarray}\label{borelpi3U}
\hat{B}\Pi_{S, \Xi}^{OPE}(p_0,T)&=-&\frac{(-1 + t)}{128\pi^4}\Bigg[ 6(1 + t)m_{s} + (-1 + t)m_{u}\Bigg]\int^{s_0(T)}_{(2m_{s}+m_{u})^2}ds \exp\Big(-\frac{s}{M^2}\Big)s^2\nonumber\\
&-&\frac{\langle\bar{q}q\rangle}{32\pi^2}(-1 +t)^2 \Bigg[ \int^{s_0(T)}_{(2m_{s}+m_{u})^2}(m_{0}^2-2s) ds\exp\Big(-\frac{s}{M^2}\Big)\Bigg]\nonumber\\
&-&\frac{3\langle\bar{s}s\rangle}{16\pi^2}(-1 + t^2)\Bigg[ \int^{s_0(T)}_{(2m_{s}+m_{u})^2} (m_{0}^2-2s)ds\exp\Big(-\frac{s}{M^2}\Big)\Bigg]\nonumber\\
&-&\frac{4\langle u\Theta^{f} u \rangle}{3\pi^2} (-1 + t^2) m_{s} p_{0}^2\nonumber\\
&+&\frac{\alpha_{s} \langle u\Theta^{g} u \rangle}{32\pi^3 N_{c}}(-1 + t)\Bigg[10(1 + t)m_{s} + m_{u} - tm_{u}\Bigg](-1 +N_{c})p_{0}^2 \nonumber\\
&+&\frac{3\langle\alpha_{s}G^2\rangle}{128\pi^3 N_{c}}(-1 + t)\Bigg[ 6(1 + t)m_{s} + m_{u}-tm_{u}\Bigg] (-1 +N_{c})\nonumber\\ &\times&\int^{s_0(T)}_{(2m_{s}+m_{u})^2}ds\exp\Big(-\frac{s}{M^2}\Big)\nonumber\\
&+&\langle\bar{s}s\rangle^2 \Bigg[\frac{3(-1 + t^2)m_{s}- (5 + 2t + 5t^2)m_{u}}{6}\Bigg]\nonumber \\
&-&\langle\bar{q}q\rangle \langle\bar{s}s\rangle \Bigg[\frac{(3 + 2t + 3t^2)m_{s} + m_{u}-t^2m_{u}}{2}\Bigg]\nonumber \\
&-&\frac{\langle\bar{q}q\rangle\langle u\Theta^{f} u \rangle}{9M^2}(-1 + t)^2(m_{0}^2-2M^2)\nonumber\\
&-&\frac{\langle\bar{s}s\rangle\langle u\Theta^{f} u \rangle}{9M^4}2(-1 + t^2) (-3m_{0}^2 M^2 + 6M^4 - 4m_{0}^2 p_{0}^2 + 16M^2 p_{0}^2)\nonumber\\
&-&\frac{\alpha_{s}\langle\bar{q}q\rangle\langle u\Theta^{g} u \rangle}{48\pi N_{c}M^4}(-1 + t)^2(-1 +N_{c}) \nonumber\\
&\times&(-2m_{0}^2 M^2+6M^4+m_{0}^2 p_{0}^2-4 M^2 p_{0}^2)\nonumber\\
&+&\frac{\alpha_{s}\langle\bar{s}s\rangle\langle u\Theta^{g} u \rangle}{24\pi N_{c}M^4}(-1 + t^2)(-1 +N_{c})\nonumber\\
&\times&(-6m_{0}^2 M^2+18M^4+5m_{0}^2 p_{0}^2-20 M^2 p_{0}^2) \nonumber\\
&+&\frac{\langle\bar{q}q\rangle\langle\alpha_{s}G^2\rangle}{64\pi N_{c}M^2}(-1 + t)^2(m_{0}^2-2M^2)(-1 +N_{c}) \nonumber\\
&-&\frac{\langle\bar{s}s\rangle\langle\alpha_{s}G^2\rangle}{32\pi N_{c}M^2}9(-1 + t^2)(m_{0}^2-2M^2)(-1 +N_{c}) \nonumber\\
&-&\frac{8\langle u\Theta^{f} u \rangle^2}{27 M^4}(-1 + t)\Bigg[(-1 + t)m_{u}(M^2 + 4p_{0}^2) \nonumber\\
&+&2(1 +t)m_{s}(3M^2 + 8p_{0}^2)\Bigg].
\end{eqnarray}

\end{document}